\documentclass[11pt]{article}
\usepackage[a4paper, margin=1in]{geometry}
\usepackage{setspace}
\usepackage{xcolor}

\usepackage{geometry}
\usepackage{amsmath, amssymb, mathtools, amsthm, mathrsfs, bbm}
\usepackage{booktabs, multirow, siunitx, pdflscape, float, threeparttable}
\usepackage{microtype, graphicx, subcaption, multirow, makecell, calc, comment, url}
\usepackage[normalem]{ulem}

\usepackage[authoryear,square]{natbib}
\usepackage{hyperref}

\newtheorem{theorem}{Theorem}
\newtheorem{lemma}{Lemma}
\newtheorem{corollary}{Corollary}
\newtheorem{assumption}{Assumption}
\newtheorem{proposition}{Proposition}
\newtheorem{remark}{Remark}
\newtheorem{example}{Example}

\newcommand\cC{\mathcal{C}}

\newcommand\cF{\mathcal{F}}
\newcommand\cG{\mathcal{G}}

\newcommand\cL{\mathcal{L}}

\newcommand\cS{\mathcal{S}}
\newcommand\cT{\mathcal{T}}

\newcommand\bN{\mathbb{N}}

\newcommand\bR{\mathbb{R}}

\newcommand\bZ{\mathbb{Z}}

\newcommand\fC{\mathfrak{C}}

\usepackage{cleveref}[capitalise]

\title{Neural Network Convergence for Variational Inequalities}
\author{Yun Zhao\thanks{Department of Mathematics, Imperial College, London SW7 2AZ, UK.
Email: yun.zhao23@imperial.ac.uk.} \ and Harry Zheng\thanks{Department of Mathematics, Imperial College, London SW7 2AZ, UK.
Email: h.zheng@imperial.ac.uk.   Supported in part by Engineering and Physical Sciences Research Council of UK  (Grant No.  EP/V008331/1).} }
\date{} 

\begin{document}

\maketitle

\begin{abstract}
 We propose an approach to applying neural networks  on linear parabolic variational inequalities. We  use loss functions that directly incorporate the variational inequality on the whole domain to bypass the need to determine the stopping region in advance and prove the existence of neural networks whose losses converge to zero. We also prove the functional convergence in the Sobolev space.  
    We then apply our approach to solving an  optimal investment and stopping  problem in finance. By leveraging duality, we convert the nonlinear HJB-type variational inequality of the primal problem  into a linear variational inequality of the dual problem and  prove the convergence of the primal value function from the dual neural network solution, an outcome made possible by our Sobolev norm analysis. We illustrate the versatility and accuracy of our method with numerical examples for both power and non-HARA utilities as well as high-dimensional American put option pricing.   Our results underscore the potential of neural networks for solving  variational inequalities in optimal stopping and control problems.
\end{abstract}

\noindent\textbf{Keywords:} Variational inequality, neural network approximation, convergence analysis, optimal investment and stopping.

\medskip
\noindent\textbf{MSC2020 Subject Classification:} 65K15, 68T07, 93E20, 60G40, 47J20.

\section{Introduction}\label{sec:one}
Deep learning has revolutionized many fields, including computer vision, natural language processing, and reinforcement learning. In partial differential equations (PDEs), deep learning has shown strong promise in solving high-dimensional problems where traditional numerical methods are computationally expensive or infeasible. Recently, deep learning to PDEs has been extended to variational inequalities, essential for optimal stopping and free-boundary problems.

Variational inequalities may be used to solve optimal stopping problems. There are also other approaches, such as martingales or Markovian chains, see \cite{peskirOptimalStoppingFreeboundary2006} for a comprehensive review. For variational inequalities, one may try to derive the free boundary, then the optimal stopping time is the first time touching, see \cite{maGlobalClosedFormApproximation2019}, or one may use the idea of Snell envelope to compare the continuation value and the immediate value, see \cite{longstaffValuingAmericanOptions2001} for a representative algorithm.

Following the approaches mentioned above, the deep learning algorithms for optimal stopping problems can be separated into three categories: 1. Approximate the solution to variational inequalities. 2. Approximate the optimal stopping time. 3. Approximate the free boundary. We next review representative works in each approach and discuss their advantages and disadvantages.

Most convergence results of deep neural networks are for PDEs. For instance, \cite{sirignanoDGMDeepLearning2018} give sufficient conditions for the functional convergence of neural networks to quasi-linear parabolic PDEs. 
They also discuss optimal stopping problems, but the loss function depends on the unknown free boundary, so they have to determine the boundary by comparing the continuation value and the immediate value, which does not follow their convergence proof for a standard PDE.
In \cite{itoNeuralNetworkBasedPolicy2021}, the authors apply policy iteration methods combined with neural networks to solve Hamilton-Jacobi-Bellman-Isaacs (HJB-I) equations and provide convergence analysis. Certain variational inequalities can be reformulated as HJB-I equations through penalization. Nevertheless, their framework is restricted to elliptic operators and relies on policy iteration, and the penalization step introduces an additional layer of convergence analysis. By contrast, our method is built directly on the minimum representation of variational inequalities, avoiding penalization altogether.
\cite{alphonseNeuralNetworkApproach2024} propose a weak adversarial neural network approach to solving variational inequalities with elliptic operators. They reformulate variational inequalities as a minimax problem and use neural networks to parametrize the solution and test functions, and then introduce a modified gradient descent ascent (GDA) algorithm to solve the optimization problem. The minimax approach ensures theoretical robustness but may require more computational resources. \cite{hureDeepBackwardSchemes2020} use neural networks to solve reflected backward stochastic differential equations (RBSDEs) related to variational inequalities, and design the algorithms by using different neural networks at every time step. Their convergence proof gives error bound in terms of the neural network approximation error and time discretization error. Our approach does not need to discretize the time variable, and only uses a single neural network for the whole domain.

Another approach is to parameterize the optimal stopping time. \cite{beckerDeepOptimalStopping2019} and \cite{beckerSolvingHighdimensionalOptimal2021} propose two deep learning algorithms to solve the optimal stopping problem in discrete time. They are based on the representation of the optimal stopping time in terms of a series of binary functions which take the values of state process at future time (backward) or past time (forward), and the value function is attained from the idea of Snell envelope. These algorithms are in discrete time and it is needed to consider the convergence from it to continuous time. One disadvantage is that they do not give the value function in the whole domain, but only at the initial time and state value, so it is inconvenient to evaluate further problems with this type of neural networks of stopping times. For instance, the derivatives of the value function would need to be approximated by finite difference method, which is often unstable with different grid sizes in practice, especially for second-order derivatives. Moreover, if there is an equation involving the first-order derivative, such as the mixed optimal stopping and control problem, it is time-consuming to evaluate the derivatives many times, making it unsuitable for these control problems.

It is also possible to approximate the free boundary directly. \cite{wangDeepLearningFree2021a} propose to parameterize both the value function and free boundary and incorporate the conditions on free boundary into the loss function. The algorithm is tested up to two dimensional Stefan problem and is proved to be effective in practice. However, there is no convergence analysis, and it is not clear if it can be generalized for higher dimensions as the dimension of the free boundary would also increase and the framework of the network needs to change accordingly. More importantly, in general, we do not know the number of free boundaries for an optimal stopping problem, nor their monotonicity or relationships with each other. For instance, in Figure 1 of \cite{maGlobalClosedFormApproximation2019}, the number of free boundaries and the continuation regions vary according to different parameters. Hence, it is difficult to estimate the free boundary for general optimal stopping problems.

In this paper, we consider a mixed optimal control and stopping problem in continuous time, see \cite{maGlobalClosedFormApproximation2019}. If we use the deep neural network for stopping times, we have to find a way to approximate the control process, and discuss the gap between discrete time and continuous time. If we use the deep neural network for free boundary, we have to know the basic properties of the unknown free boundary and it is difficult to be generalized to high dimensions. The value function  is a solution to a nonlinear variational inequality that is typically difficult to solve. We turn to the dual method and formulate a dual optimal stopping problem.  The dual value function  satisfies a linear variational inequality that can be solved with deep neural networks, and the primal value function can then be recovered with the primal-dual relation. 
This approach depends crucially on the accuracy of the neural network approximation to the solution of the linear variational inequality.

There are few results for convergence proof of deep neural networks to variational inequalities in the literature. We design an algorithm that does not require a prior knowledge of the free-boundary and converges to the solution. We also solve the optimal investment stopping problem, which results in a fully nonlinear variational inequality, with the dual method and maintain the convergence for the primal value function after recovering it from the dual relation.  We require $H^{0,1}$ convergence  of the neural network to the dual value function, more than $L^2$ convergence,  to ensure the convergence to the primal value function in $C^0$ at almost any time with Sobolev embedding theorem. To prove such strong convergence, we use an idea from functional analysis: right inverse of trace operator. There are many types of trace theorems that ensure a continuous right inverse map.
In general, there is a gain in interior regularity from boundary regularity, see Theorems 1.5.1.1-1.5.1.3 in \cite{grisvardEllipticProblemsNonsmooth1985} for both trace and inverse trace theorems in the elliptic case. There are also trace theorems in parabolic case, see Theorem 4.2.3 in \cite{lionsNonHomogeneousBoundaryValue1972a}. By incorporating the fractional Sobolev norm (also called Sobolev-Slobodeckij norm) on the boundary into the loss function, we obtain a higher order  convergence in the interior, and overcome the technical difficulties in the convergence analysis. A similar type of norm also appears in \cite{itoNeuralNetworkBasedPolicy2021} for establishing convergence of elliptic HJB-I equations, where the argument relies not on an inverse trace theorem but instead on an $H^2$ estimate for the Dirichlet problem.

Another challenge in the convergence proof is when we use numerical methods to solve variational inequalities, the minimum between two approximations typically have kink points, where the second-order derivative is a Dirac delta distribution. The traditional theories about variational inequalities do not cover this case. We overcome this difficulty by generalizing the results from \cite{dautovPenaltyMethodsOneSided2021} for piecewise $H^2$ obstacles.

The main contributions of this paper are as follows. Our work is unique because we directly incorporate the minimum representation of variational inequalities into the loss function and prove the convergence, so there is no need to determine the stopping region in advance. We combine the trace theorem 4.2.3 in \cite{lionsNonHomogeneousBoundaryValue1972a} with Bartle-Graves theorem (see Theorem 5.1 in \cite{messerschmidtPointwiseLipschitzSelection2019}) to get an inverse trace theorem in the parabolic case for convergence analysis, which is innovative and still admits the universal approximation theorem. We also overcome the difficulties when the obstacles are in the minimum form. Moreover, we extend the convergence results to unbounded domains and prove the existence of a single sequence of neural networks that converges to the solution in any compact set, which is the case in most optimal stopping problems. Finally, the strong convergence in Sobolev space allows us to prove the uniform convergence for a HJB type of nonlinear variational inequality arising from mixed optimal stopping and control problems. We use deep neural networks to approximate the dual value function and prove it converges to the primal value function after a transformation. Numerical tests show that our method is more efficient and accurate than traditional methods, see Example \ref{exa:1}. We also test our method on American put option pricing in high dimensions, see Example \ref{exa:2}.  Comparing with the results from \cite{hureDeepBackwardSchemes2020}, which originates from BSDE and requires multiple neural networks at different time steps for  Bermudan options, we get similar-level accuracy by using only a single neural network.

The remainder of the paper is organized as follows.  In Section \ref{sec:two}, we state the design of the deep neural network and prove the existence of the neural networks that can drive the loss function to zero, see Theorem \ref{thm:loss_function_convergence}. In Section \ref{sec:three}, we use the inverse trace theorem and penalty method to prove that the sequence of neural networks from Theorem \ref{thm:loss_function_convergence} indeed converges to the real solution in the Sobolev space, see Theorems \ref{thm:f_n_converges_to_u} and \ref{thm:g_n_to_u_in_unbounded_domain}. In Section \ref{sec:four}, we use the dual method to solve a nonlinear variational inequality and show the convergence to the primal value function, see Theorem \ref{thm:uniform_convergence_to_primal_value}. In Section \ref{sec:five}, we illustrate our method with numerical examples  and compare it with traditional methods. Section  \ref{app:proofs} contains the proofs of all theorems.
 Section \ref{sec:conclusions} concludes. 

We next list some notations and spaces used in the paper.

{\bf Notations}. Denote by $T>0$ a given terminal time, $\Omega \subset \bR^d$ a bounded open set, $\Omega_T := (0,T) \times \Omega$,  $\overline{\Omega_T}:=[0,T]\times\overline{\Omega}$ the closure of $\Omega_T$,  $\Gamma:=\partial\Omega$ the spatial boundary  which is required to be smooth (for the trace theorem and Rellich-Kondrachov theorem),   $\Sigma:=(0,T)\times\Gamma$  the lateral boundary, and  $\partial_p\Omega_T:= (\{0\}\times\overline{\Omega})\cup \Sigma$ the parabolic boundary of $\Omega_T$.  
For a function $u: \Omega_T \rightarrow \bR$, denote by $\partial_t^k u = \partial^k u / \partial t^k$, $\partial_{x_i}^k u = \partial^k u / \partial x_i^k$ for $k \geq 1$, $\partial_{x_i x_j} u = \partial^2 u / \partial x_i \partial x_j$, etc., 
$\nabla u= \left(\partial_{x_1}u, \dots, \partial_{x_d}u\right)$ the gradient, and
\[D^{\alpha}_x u(t,x) \;=\; \frac{\partial^{\alpha_1 + \cdots + \alpha_d} u}{\partial x_1^{\alpha_1} \cdots \partial x_d^{\alpha_d}}(t,x),\]
where   $\alpha = (\alpha_1, \alpha_2, ..., \alpha_d)$ of nonnegative integers is  a multi-index with $|\alpha| := \sum_{i=1}^d \alpha_i$. 
The use of weak derivatives 
\footnote{
    A function $v \in L^1_{\text{loc}}(\Omega)$ is called the \emph{weak derivative} of $u$ with respect to variable $x_i$ if
  $
    \int_{\Omega} u(x) \, \partial \varphi(x) / \partial x_i \, dx = -\int_{\Omega} v(x) \, \varphi(x) \, dx
    $
    for every test function $\varphi \in C_c^\infty(\Omega)$ (i.e., $\varphi$ is infinitely differentiable with compact support in $\Omega$).
    }%
will be specified when necessary. 
We need to introduce the tangential derivative on the boundary $\Gamma$. At each $x\in\Gamma$, choose an orthonormal basis $\{\tau_1(x), \cdots, \tau_{d-1}(x)\}$ of the tangent space $T_x\Gamma$. For a multi-index $\beta = (\beta_1 ,\dots, \beta_{d-1})$, set
\[
D_\tau^\beta u(t,x) \;=\; (\tau_1\cdot\nabla)^{\beta_1}\cdots(\tau_{n-1}\cdot\nabla)^{\beta_{n-1}} u(t,x),
\]
where $(\tau_1\cdot\nabla)^{\beta_i}u(t,x) = \frac{\partial^{\beta_i}}{\partial s^{\beta_i}} u(t,x+s\tau_i(x)) |_{s=0}$ for $ (t,x) \in [0,T] \times \Gamma$. 

{\bf H\"older spaces}. 
Following \cite{friedmanVariationalPrinciplesFreeboundary1982}, define $C^\alpha(\overline{\Omega}_T)$ as the space of all H\"older continuous functions with exponent $\alpha \in (0,1)$, i.e.,
\[
H_\alpha(u) \;=\; \sup_{(t,x), (t',x') \in \Omega_T} \frac{|u(t,x) - u(t',x')|}{(|t-t'|^{1/2} + |x-x'|)^\alpha} < \infty.
\]
$C^\alpha(\overline{\Omega}_T)$ is a Banach space with  norm $\|u\|_\alpha \;=\; \sup_{(t,x) \in {\Omega}_T} |u(t,x)| + H_\alpha(u)$.

{\bf Sobolev spaces with integer exponents}.
For integers $s, r \geq 0$, with weak derivative, define the Sobolev space $W_p^{s,r}(\Omega_T)$ as
\[
    W_p^{s,r}(\Omega_T) := \left\{ u \in L^p(\Omega_T) \,\middle|\, \partial_t^k u, D_x^\alpha u \in L^p(\Omega_T) \text{ for } k \leq s, |\alpha| \leq r \right\},
\]
and $H^{s,r}(\Omega_T)=W_2^{s,r}(\Omega_T)$ with norm $
\|u\|_{H^{s,r}(\Omega_T)} = ( \sum_{k\leq s} \|\partial_t^k u\|_{L^2(\Omega_T)}^2 + \sum_{|\alpha|\leq r} \|D_x^\alpha u\|_{L^2(\Omega_T)}^2)^{1/2}$.
If $s=r$, we denote $H^{s,s}(\Omega_T)$ as $H^{s}(\Omega_T)$. When there is no time variable, we denote 
\[
W^{r,p}(\Omega) := \left\{ u \in L^p(\Omega) \, \middle| \, D_x^\alpha u \in L^p(\Omega) \text{ for  }  |\alpha| \leq r \right\},
\]
and $H^r(\Omega)=W^{r,2}(\Omega)$ with  norm $
\|u\|_{H^r(\Omega)} = ( \sum_{|\alpha|\leq r} \|D_x^\alpha u\|_{L^2(\Omega)}^2 )^{1/2}$.
We use $H^1_0(\Omega)$ to denote the functions in $H^1(\Omega)$ that vanishes on the boundary $\partial\Omega$. The spaces $W_p^{s,r}(\Sigma)$, $H^{s,r}(\Sigma)$, $H^s(\Sigma)$ for the lateral boundary $\Sigma$, and $W^{r,p}(\Gamma)$, $H^r(\Gamma)$ for the space boundary $\Gamma$ are similarly defined, using the tangential derivatives $D_\tau^\beta u$. The definition is independent of the particular choice of local frames $\{\tau_1(x), \cdots, \tau_{d-1}(x)\}$ at each $x\in\Gamma$, see e.g. \cite{lionsNonHomogeneousBoundaryValue1972}.

{\bf Sobolev spaces with non-integer exponents}.
In the following, we generalize the integer exponents to non-integers and introduce the fractional Sobolev space, also called Sobolev-Slobodeckij space. We would use it for applying the trace theorem, which is essential in our convergence analysis. For more details, see \cite{brezisFunctionalAnalysisSobolev2011}, \cite{lunardiAnalyticSemigroupsOptimal1995}, \cite{lionsNonHomogeneousBoundaryValue1972a} and \cite{demengelFunctionalSpacesTheory2012}.
For $s,r>0$ and $s,r$ are not an integer, denote $\rho=s-\lfloor s\rfloor\in(0,1)$ and $\theta=r-\lfloor r\rfloor\in(0,1)$, and define the Sobolev-Slobodeckij norms for $p=2$ as
\[
[f]_{\rho,(0,T)}:=\left(\int_0^T\int_0^T\frac{\|f(t_1,\cdot)-f(t_2,\cdot)\|_{L^2(\Gamma)}^2}{|t_1-t_2|^{2\rho+1}} \, dt_1 \, dt_2 \right)^{1/2},
\]
\[
[f]_{\theta,\Gamma}:=\left(\int_\Gamma\int_\Gamma\frac{|f(x)-f(y)|^2}{\|x-y\|^{2\theta+d-1}} \, d\sigma(x) \, d\sigma(y) \right)^{1/2},
\]
where $d\sigma(x)$ is the surface measure on $\Gamma$.
The norm in $H^{r,s}(\Sigma)$ is defined by $\|f\|_{H^{r,s}(\Sigma)}:=(\|f\|^2_{H^{\lfloor r\rfloor, \lfloor s\rfloor}(\Sigma)}+[f]^2_{\rho,(0,T)}+\int_0^T[f(t,\cdot)]^2_{\theta,\Gamma}\,dt)^{1/2}$ and that in $H^r(\Gamma)$ by $\|f\|_{H^r(\Gamma)}:=(\|f\|^2_{H^{\lfloor r\rfloor}(\Gamma)}+[f]^2_{\theta,\Gamma})^{1/2}$.

\section{Model Formulation and Neural Network Approximations} \label{sec:two}
Consider the following linear variational inequality with boundary conditions:
\begin{align}
    \label{eq:original_VI}
    \begin{cases}
        \min\{\cG[u](t,x), u(t,x)-g(t,x)\} = 0, & \text{for } (t,x) \in \Omega_T, \\
        u(t,x) = g(t,x), & \text{for } (t,x) \in \partial_p \Omega_T,
    \end{cases}
\end{align}
where $\partial_p\Omega_T$ is the parabolic boundary of $\Omega_T$ and  $\cG$ is a linear parabolic operator of the form 
\begin{equation} \label{eq:operator_G}
\mathcal{G}[u](t,x) := \partial_t u - \sum_{i,j=1}^d a_{ij}(t,x) \partial_{x_i x_j} u + \sum_{i=1}^d b_i(t,x) \partial_{x_i} u + c(t,x) u.
\end{equation}
The bilinear form associated with $\cG$ at a fixed time $t$ is
\begin{align} \label{eq:bilinear_form}
a(t; u, v) := \int_\Omega \left(\sum_{i,j=1}^d a_{ij}(t,x) \, \partial_{x_i} u \, \partial_{x_j} v + \sum_{i=1}^d (b_i(t,x) + \sum_{j=1}^d \partial_{x_j} a_{ij}(t,x)) \partial_{x_i} u \, v + c(t,x) u v\right) dx,
\end{align}
which can be understood as the result after testing $\cG[u]$ with $v$ that vanishes on the boundary and using integration by parts.
Define the convex set
\[
K := \{v \in H^1(\Omega_T) \,|\, v=g \text{ on } \partial_p\Omega_T, v \ge g \text{ a.e.}\}.
\]
\eqref{eq:original_VI} is equivalent to the following problem (see \cite{friedmanVariationalPrinciplesFreeboundary1982}):  Find $u\in K$ such that  
\[
\int_\Omega \partial_t u \, (v-u) \, dx + a(t; u, v-u) \geq 0, \quad \text{ for a.a. } t \in (0,T), \, \forall v \in K.
\]

Define the stopping region as $\cS := \{(t,x) \in \Omega_T \, | \, u(t,x) = g(t,x)\}$ and the continuation region as $\cC := \{(t,x) \in \Omega_T \, | \, \cG[u](t,x) = 0, u(t,x) > g(t,x)\}$. These regions do not overlap and $\cC \cup \cS = \Omega_T$. The following assumption ensures the appropriate properties of the solution to \eqref{eq:original_VI}, see \cite{friedmanVariationalPrinciplesFreeboundary1982}.
\begin{assumption}[Adapted from {\cite{friedmanVariationalPrinciplesFreeboundary1982}}]
    \label{assum:assumptions_on_Theorem_8.2}
    \mbox{}
    \begin{itemize}
    \item (uniformly parabolic) There exists a constant $\lambda > 0$ independent of $(t,x)$ such that 
    $\sum_{i,j=1}^d a_{ij}(t,x) \xi_i \xi_j \geq \lambda |\xi|^2$ for $\xi \in \bR^d$. 
     \item (coercivity) There exists a constant $M > 0$ such that $a(t; u, u) \geq  M \|u\|_{H^1(\Omega)}^2$ for $u \in H^1(\Omega)$.
        \item The coefficients satisfy 
    $\sum_{i,j=1}^d \|a_{ij}\|_{\alpha} + \sum_{i=1}^d \|b_i\|_\alpha + \|c\|_\alpha \leq K$
    for some constant $K$ and $\alpha \in (0,1)$, where $\|\cdot\|_{\alpha}$ is the H\"older norm.
    \item $c(t,x) \geq 0$, $\partial\Omega\in C^{2+\alpha}$ \footnote{An open set $\Omega \subset \mathbb{R}^d$ is said to have $C^{2+\alpha}$ boundary if, 
    for every point $x_0 \in \partial\Omega$, there exist an $r > 0$ and a 
    $C^{2+\alpha}$ function 
  $    \gamma : \mathbb{R}^{d-1} \;\to\; \mathbb{R}$
    such that, after possibly relabeling and reorienting the coordinate axes, we have
    $
    \Omega \cap B(x_0,r)
    =
    \left\{x \in B(x_0, r): x_d > \gamma(x_1,\dots, x_{d-1}) \right\} \text{ and }
    \partial\Omega \cap B(x_0,r)
    =
    \left\{x \in B(x_0,r): x_d = \gamma(x_1,\dots,x_{d-1})\right\}.
    $
    Similarly, we can define the set with Lipschitz boundary with $\gamma$ being Lipschitz continuous, see \cite{leoniFirstCourseSobolev2009}.}.
    \item $g, \partial_t g, D_x g, D_x^2 g$ belong to $C^\alpha(\overline{\Omega}_T)$.
    \item $|\nabla a_{ij}| \leq C$ for some constant $C$ and all $i,j=1,...,d$.
    \end{itemize}
\end{assumption}
Under Assumption \ref{assum:assumptions_on_Theorem_8.2}, there is a unique classical solution to \eqref{eq:original_VI} in $C(\overline{\Omega}_T) \cap W_p^{1,2}(\Omega_T)$ for any $1 < p < +\infty$, see Theorem 1.8.2 in \cite{friedmanVariationalPrinciplesFreeboundary1982}.

We next introduce  a set of neural networks with one hidden layer, $n$ nodes and activation function $\psi$:
\begin{equation}
    \label{def:fC^n}
    \fC^n(\psi) := \left\{ \zeta(t,x): \bR^{1+d} \rightarrow \bR \left| \zeta(t,x) = \sum_{i=1}^n \beta_i \psi \left( \alpha_{0,i} t + \sum_{j=1}^d \alpha_{j,i} x_j + c_i \right)\right.
    \right\},
\end{equation}
and parameters $\theta := \{\beta_1,..., \beta_n, \alpha_{0,1}, ..., \alpha_{d,n}, c_1, ..., c_n\} \in \bR^{2n+n(1+d)}$. Define the set of all neural networks with one hidden layer and activation function $\psi$ as
\[
\fC(\psi) := \bigcup_{n \geq 1} \fC^n(\psi).
\]

For a neural network $f(t,x;\theta)\in\fC(\psi),$ define the difference between it and the solution on the initial-time boundary
\[e(x;\theta):=f(0,x;\theta)-g(0,x) \text{ for } x\in \Omega,\]
and the difference on the lateral boundary
\[d(t,x;\theta):=f(t,x;\theta)-g(t,x) \text{ for } (t,x) \in \Sigma,\]
and its normal derivative on the lateral boundary
\[d_1(t,x;\theta):=\frac{\partial}{\partial\nu} (f(t,x;\theta)-g(t,x)) = \nabla(f(t,x;\theta)-g(t,x)) \cdot \nu(x) \text{ for } (t,x) \in \Sigma,\]
where $\nu(x)$ is the unit normal vector pointing outward from $\Omega$.

Define the following loss function
\begin{equation} \label{def:J_f}
    J(f) = \|\min\{\cG[f], f-g\}\|^2_{L^2(\Omega_T), \mu_1} + \|e\|^2_{H^{1}(\Omega), \mu_2} + \|d\|^2_{H^{\frac{3}{4},\frac{3}{2}}(\Sigma),\mu_3} + \|d_1\|^2_{H^{\frac{1}{4},\frac{1}{2}}(\Sigma),\mu_3},
\end{equation}
where $\mu_1, \mu_2, \mu_3$ are the probability measures on $\Omega_T, \Omega, \Sigma$ respectively. We require them to be absolutely continuous with respect to the Lebesgue measure and bounded away from zero. In practice, we can use uniform or truncated normal distribution.

\begin{remark}
    The fractional Sobolev norms allow us to have a better convergence for $f(t,x;\theta)$ on the boundary, and we can then use the inverse trace theorem to construct auxiliary functions of $f(t,x;\theta)$. The simplest example for the linear inverse trace operator is from $H^{\frac{1}{2}}(\partial\Omega)$ to $H^1(\Omega)$. In this paper, we use the trace Theorem 4.2.3 in \cite{lionsNonHomogeneousBoundaryValue1972a}, which gives a linear inverse mapping from $H^{1}(\Omega)$, $H^{\frac{3}{4},\frac{3}{2}}(\Sigma)$ and $H^{\frac{1}{4},\frac{1}{2}}(\Sigma)$ to $H^{1,2}(\Omega_T)$.
\end{remark}

The following proposition shows that $\fC(\psi)$ is uniformly $2$-dense on compact sets in $C^2(\bR^{1+d})$.
\begin{proposition}[Universal Approximation Theorem, \cite{hornikApproximationCapabilitiesMultilayer1991}]
    \label{prop:universal_approximation_theorem}
    If the activation function $\psi \in C^2(\bR^{1+d})$ is nonconstant and bounded, then $\fC(\psi)$ is uniformly $2$-dense on compact sets in $C^2(\bR^{1+d})$, which means that for every compact set \(K \subset \mathbb{R}^{1+d}\), every function \(\phi \in C^2(\mathbb{R}^{1+d})\), and every \(\epsilon > 0\), there exists \(f \in \fC(\psi)\) such that
    \[
    \max_{|\alpha|\leq 2}\sup_{z\in K} |D^\alpha (f(z)-\phi(z))| < \epsilon.
    \]
\end{proposition}
Following the same reasoning as \cite{sirignanoDGMDeepLearning2018}, for any $u \in C^{1,2}([0,T]\times \bR^d)$ and $\epsilon > 0$, one can find $f \in \fC(\psi)$ such that
\begin{equation}
    \label{eq:2-dense_property}
    \max_{\alpha\leq1, \alpha+|\beta| \leq 2} \sup_{(t,x) \in \overline{\Omega}_T} |\partial_t^\alpha D_x^{\beta} u(t,x) - \partial_t^\alpha D_x^{\beta} f(t,x;\theta)| < \epsilon.
\end{equation}

Since we only have $u \in C(\overline{\Omega}_T)\cap W_p^{1,2}(\Omega_T)$, to make use of the universal approximation theorem, we need the following assumption.
\begin{assumption}
    \label{assum:conditions_on_u}
The problem \eqref{eq:original_VI} admits a solution $u \in C^{1,2}(\overline{\Omega}_T)$, where the value of $u$ and its derivatives at the boundary are defined as one-sided limits, see Chapter 4 in \cite{rudinPrinciplesMathematicalAnalysis1976}.
\end{assumption}
Under Assumption \ref{assum:conditions_on_u}, we use the Whitney extension theorem (see \cite{Whitney1992}) to extend $u$ to $C^{1,2}([0,T]\times\bR^d)$ and then apply the universal approximation theorem to get a network $f \in \fC(\psi)$ that satisfies \eqref{eq:2-dense_property}.

\begin{theorem}
    \label{thm:loss_function_convergence}
Let Assumptions \ref{assum:assumptions_on_Theorem_8.2} and \ref{assum:conditions_on_u} hold. If the activation function $\psi \in C^2(\bR^{1+d})$ is nonconstant and bounded, then for any $\epsilon > 0$, there exists  $f \in \fC(\psi)$ such that
$J(f) < \epsilon$.

\begin{proof}
    See \ref{app:proof_thm_1}.
\end{proof}
\end{theorem}

\section{Strong Convergence of Neural Network Approximations} \label{sec:three}
In this section, we show that if a sequence of neural networks drives the loss function to zero, then it converges to the solution $u$ strongly in $H^{0,1}$. We first establish the convergence in the bounded domain $\Omega_T$ and then extend the result to the unbounded domain $(0,T) \times \bR^d$.

\subsection{Convergence in Bounded Domain}

Theorem \ref{thm:loss_function_convergence} gives sufficient conditions for the existence of $\{f(t,x;\theta_n)\}_{n=1}^\infty$, which we denote as $\{f_n(t,x)\}_{n=1}^\infty$, in $\fC(\psi)$ such that $ J(f_n) \to 0$ as $n \to +\infty$. Denote by $w_n := \min\{\cG[f_n], f_n-g\}$ on $\Omega_T$ and $v_n := f_n - g$ on $\partial_p \Omega_T$. The corresponding equations are
\begin{align}
    \label{eq:VI_f_n}
    \begin{cases}
        \min\{\cG[f_n](t,x), f_n(t,x)-g(t,x)\} = w_n(t,x), & \text{for } (t,x) \in \Omega_T, \\
        f_n(t,x) = g(t,x) + v_n(t,x), & \text{for } (t,x) \in \partial_p \Omega_T,
    \end{cases}
\end{align}
where
\[
\|w_n\|_{L^2(\Omega_T), \mu_1}, \, \|v_n(0, \cdot)\|_{H^1(\Omega), \mu_2}, \, \|v_n\|_{H^{\frac{3}{4}, \frac{3}{2}}(\Sigma),\mu_3}, \, \|\frac{\partial}{\partial\nu}v_n\|_{H^{\frac{1}{4},\frac{1}{2}}(\Sigma),\mu_3} \to 0
\] 
as $n \to +\infty$. Since $\mu_1, \, \mu_2, \, \mu_3$ are absolute continuous with respect to the Lebesgue measure and bounded away from zero, we have
\begin{equation} \label{eq:convergence_of_w_n_and_v_n}
\|w_n\|_{L^2(\Omega_T)}, \, \|v_n(0, \cdot)\|_{H^1(\Omega)}, \, \|v_n\|_{H^{\frac{3}{4}, \frac{3}{2}}(\Sigma)}, \, \|\frac{\partial}{\partial\nu}v_n\|_{H^{\frac{1}{4},\frac{1}{2}}(\Sigma)} \to 0
\end{equation}
as $n \to +\infty$. $w_n$ and $v_n$ are originally defined on $\Omega_T$ and $\partial_p\Omega_T$ respectively. $w_n$ can be extended to $\partial_p\Omega_T$ naturally by its continuity. However, it is not trivial to extend $v_n$ to $\Omega_T$. We need to ensure after extension, $v_n$ converges to zero in $H^1(\Omega_T)$ and is uniformly bounded in $H^{1,2}(\Omega_T)$ if we want to prove strong convergence in $H^{0,1}(\Omega_T)$. This turns out to be an application of the inverse trace theorems, which requires the  loss function to include the fractional Sobolev norms.
\begin{theorem}
    \label{thm:inverse_trace_theorem}
    (Inverse Trace Theorem, combining Theorem 4.2.3 in \cite{lionsNonHomogeneousBoundaryValue1972a} and Bartle-Graves theorem (see Theorem 5.1 in \cite{messerschmidtPointwiseLipschitzSelection2019})) Let $\Omega$ be a bounded open set in $\bR^d$ with $C^{2+\alpha}$ boundary $\Gamma$ for some $\alpha \in (0,1)$ and $T>0$. For any $e \in H^{1}(\Omega)$, $d \in H^{\frac{3}{4},\frac{3}{2}}(\Sigma)$ and $d_1 \in H^{\frac{1}{4},\frac{1}{2}}(\Sigma)$ satisfying the compatibility condition $e|_{\partial\Omega} = d|_{t=0}$, there exists a function $v \in H^{1,2}(\Omega_T)$ such that
    \[v(0,\cdot)|_{\Omega} = e, \quad v|_\Sigma = d, \quad \frac{\partial v}{\partial \nu}\bigg|_\Sigma = d_1,\]
    and
    \[\|v\|_{H^{1,2}(\Omega_T)} \leq C(\|e\|_{H^{1}(\Omega)} + \|d\|_{H^{\frac{3}{4},\frac{3}{2}}(\Sigma)} + \|d_1\|_{H^{\frac{1}{4},\frac{1}{2}}(\Sigma)}),\]
    where $C$ is a constant depending only on $\Omega$ and $T$.
    \begin{proof}
        By setting $r=2$ and $s=1$ in Theorem 4.2.3 in \cite{lionsNonHomogeneousBoundaryValue1972a}, noting that the compatibility conditions are satisfied with these choices of $r$ and $s$, we know that
        there exists a continuous linear surjective mapping from $H^{1,2}(\Omega_T)$ to $H^{1}(\Omega) \times H^{\frac{3}{4},\frac{3}{2}}(\Sigma) \times H^{\frac{1}{4},\frac{1}{2}}(\Sigma)$. The half-space and infinite time in Theorem 4.2.3 are only special cases, and it can be extended to our case where $\Omega$ is bounded and $T<\infty$ as  fractional Sobolev spaces are local, see Section 1.7 in \cite{lionsNonHomogeneousBoundaryValue1972} and Theorem 4.2.1 in \cite{lionsNonHomogeneousBoundaryValue1972a}. By Bartle-Graves theorem (see Theorem 5.1 in \cite{messerschmidtPointwiseLipschitzSelection2019}), there exists a continuous right inverse of the trace operator, which is the desired inverse trace operator.
    \end{proof}
\end{theorem}
\begin{remark}
    The similar results of inverse trace theorem also hold in the elliptic case, see Theorems 1.3.2 and 1.4.2 in \cite{lionsNonHomogeneousBoundaryValue1972}, where the surjectivity of the trace operator is proved by constructing an extension which is a multiplication of the trace and guarantees the continuous inverse mapping. Theorem 1.5.1.2 in \cite{grisvardEllipticProblemsNonsmooth1985} also gives a similar result for the elliptic case.
\end{remark}
Since $v_n=f_n-g$ on $\partial_p\Omega_T$, the compatibility condition in Theorem \ref{thm:inverse_trace_theorem} is satisfied. By \eqref{eq:convergence_of_w_n_and_v_n}, we have an extension $v_n \to 0$ in $H^{1,2}(\Omega_T)$. For simplicity, we still denote the extension as $v_n$.

\begin{remark}
    One might want to use a smooth function to mollify $w_n$ and the mollified function would converge to zero in $H^{r}(\Omega_T)$ for any $r\geq 0$. However, there must be compatible conditions on the source and boundary terms of the variational inequality, i.e., $w_n \geq v_n$. If we mollify $w_n$, the mollified function might be less than $v_n$.
\end{remark}

Next, consider the auxiliary variational inequality with $v_n$ as the source term and denote the solution as $\hat{f}_n$:
\begin{align}
    \label{eq:auxilliary_VI}
    \begin{cases}
        \min\{\cG[\hat{f}_n](t,x), \hat{f}_n(t,x)-g(t,x)\} = v_n(t,x), & \text{for } (t,x) \in \Omega_T, \\
        \hat{f}_n(t,x) = g(t,x) + v_n(t,x), & \text{for } (t,x) \in \partial_p \Omega_T.
    \end{cases}
\end{align}

In this section, we prove $f_n \to u$ strongly in $H^{0,1}(\Omega_T)$. We decompose it into two parts: In Theorem \ref{thm:hat_f_n_converges_to_u}, we prove $\hat{f}_n \to u$ strongly in $H^{0,1}(\Omega_T)$. In Theorem \ref{thm:f_n_converges_to_hat_f_n}, we prove $\hat{f}_n - f_n \to 0$ strongly in $H^{0,1}(\Omega_T)$. After combining them, we get $f_n \to u$ strongly in $H^{0,1}(\Omega_T)$ in Theorem \ref{thm:f_n_converges_to_u}.

\begin{theorem}
    \label{thm:hat_f_n_converges_to_u}
Let Assumption \ref{assum:assumptions_on_Theorem_8.2} hold. If there is a sequence $\{f_n\}_{n=1}^\infty$ such that $J(f_n)\to0$, then the corresponding auxiliary functions $\hat{f}_n \to u$ strongly in $H^{0,1}(\Omega_T)$, where $u$ is the unique solution to \eqref{eq:original_VI}.

\begin{proof}
    See \ref{app:proof_thm_2}.
\end{proof}
\end{theorem}

Next, we compare \eqref{eq:VI_f_n} and \eqref{eq:auxilliary_VI} using a prevalent technique in variational inequalities, the penalty method, see \cite{friedmanVariationalPrinciplesFreeboundary1982} and \cite{bensoussanApplicationsVariationalInequalities2011}. In classical theories about strong convergence in $H^{0,1}(\Omega_T)$, it is required that the obstacle, here $g+w_n$ in \eqref{eq:VI_f_n} and $g+v_n$ in \eqref{eq:auxilliary_VI}, should be at least in $H^{1,2}(\Omega_T)$. A standard version of the proof can be found in Proposition 2.1 of \cite{adamsOptimalControlObstacle2002} and Proposition 5.2.1 in \cite{barbuAnalysisControlNonlinear1993}. In our case, the main difficulty is that the obstacle $g+w_n$ might have kink points where the second-order derivatives exist as a delta distribution, because $w_n$ is the minimum between $\cG[f_n]$ and $f_n - g$. Therefore, $w_n$ is not in $H^{1,2}(\Omega_T)$ in general. To address this problem, we generalize an existing result from \cite{dautovPenaltyMethodsOneSided2021}.

Note that $w_n(t,\cdot)$ is piecewise $H^2$ in $\Omega$ at any time $t \in (0,T)$, i.e., there exists a partition of $\Omega$, $\{\Omega_{1,n}^t, \Omega_{2,n}^t, \dots, \Omega_{m_t^n,n}^t\}$, $m_t^n \geq 1$ such that each subdomain $\Omega_{i,n}^t$ has a Lipschitz boundary $\partial\Omega_{i,n}^t$,
    \[
    \Omega_{i,n}^t \cap \Omega_{j,n}^t = \emptyset, \quad \forall i \neq j, \quad \bigcup_{i=1}^{m_t^n} \overline{\Omega_{i,n}^t} = \overline{\Omega},
    \]
    and
    \[
    w_n(t,\cdot)|_{\Omega_{i,n}^t} \in H^2(\Omega_{i,n}^t), \quad 1\leq i \leq m_t^n, \quad \forall t \in (0,T).
    \]
    Denote by $\{\Gamma_{1,n}^t, \Gamma_{2,n}^t, \dots, \Gamma_{M_t^n,n}^t\}$ all nonempty intersections $\partial\Omega_{i,n}^t \cap \partial\Omega_{j,n}^t$ for $1\leq i \neq j \leq m_t^n$. For $i=1,\dots, m_t^n$, $\partial\Omega_{i,n}^t$ is closed because $(\partial\Omega_{i,n}^t)^c = (\Omega_{i,n}^t)^\circ \cup ((\Omega_{i,n}^t)^c)^\circ$ is open. Therefore, $\Gamma_{i,n}^t$ is closed, $i=1,\dots,M_t^n$. Let $\Gamma_n^t:=\bigcup_{k=1}^{M_t^n} \Gamma_{k,n}^t$. $\Gamma_n^t$ is also closed.

\begin{assumption}
    \label{assum:w_n}
    \mbox{}
    \begin{itemize}
    \item $w_n$ is uniformly bounded in $H^1(\Omega_T)$.
    \item $\sum_{i=1}^{m_t^n} \|w_n(t,\cdot)\|_{H^2(\Omega_{i,n}^t)}^2: (0,T) \to \bR$ is measurable in $t$.
    \item $\sup_{n} \int_0^T \sum_{i=1}^{m_t^n} \|w_n(t,\cdot)\|_{H^2(\Omega_{i,n}^t)}^2 \, dt < \infty$.
    \item The multifunction $t \mapsto \Gamma_n^t \subset \Omega$ has a measurable graph in $\Omega_T$, see Theorem 8.1.4 in \cite{aubinSetValuedAnalysis2009} for related discussions.
    \item The boundaries $\partial\Omega_{i,n}^t$ are uniformly Lipschitz.
    \end{itemize}
\end{assumption}

\begin{remark}
    When analyzing deep neural networks, people often make some assumptions on the uniform bound because little is known about how they decrease the loss function. For example, Theorem 7.3 in \cite{sirignanoDGMDeepLearning2018} assumes the neural networks are uniformly bounded in $L^2(\Omega_T)$ to prove the functional convergence.
\end{remark}

\begin{remark}
    \label{rem:weak_derivative_of_min}
    The kink points of $w_n$ can only happen on the set $\{(t,x): \cG[f_n](t,x) = f_n(t,x) - g(t,x), \nabla \cG[f_n](t,x) \neq \nabla (f_n(t,x) - g(t,x))\}$. From Comments on Chapter 9 in \cite{brezisFunctionalAnalysisSobolev2011}, since $f_n$ is designed by an activation function in $C^{2}(\bR^{d+1})$, and $g$ is under Assumption \ref{assum:assumptions_on_Theorem_8.2}, the Lebesgue measure of this set is zero. Hence, the first-order weak derivative of $w_n$ exists in $\Omega_T$.
\end{remark}

\begin{assumption}
    \label{assum:w_n_equal_v_n}
    $w_n(t,x) = v_n(t,x)$ a.e. for $(t,x) \in \partial_p\Omega_T$.
\end{assumption}
\begin{remark}
    Assumption \ref{assum:w_n_equal_v_n} is reasonable because according to \eqref{eq:original_VI}, $u-g \leq \cG[u]$ on $\partial_p\Omega_T$. Therefore, from the universal approximation theory of the neural networks, it is resonable to expect that $f_n-g \leq \cG[f_n]$ on $\partial_p\Omega_T$, which means in \eqref{eq:VI_f_n}, $w_n = f_n - g = v_n$ on $\partial_p\Omega_T$.
\end{remark}

\begin{theorem}
\label{thm:f_n_converges_to_hat_f_n}
Let Assumption \ref{assum:assumptions_on_Theorem_8.2} hold. Suppose there exists a sequence $\{f_n\}_{n=1}^\infty$ such that $J(f_n)\to0$. If Assumptions \ref{assum:w_n} and \ref{assum:w_n_equal_v_n} hold, then $\hat{f}_n - f_n\to 0$ strongly in $H^{0,1}(\Omega_T)$.
\begin{proof}
    See \ref{app:proof_thm_3}.
\end{proof}
\end{theorem}
Combining Theorems \ref{thm:loss_function_convergence}, \ref{thm:hat_f_n_converges_to_u} and \ref{thm:f_n_converges_to_hat_f_n}, we have the following convergence result.
\begin{theorem}
    \label{thm:f_n_converges_to_u}
    Let Assumptions \ref{assum:assumptions_on_Theorem_8.2} to \ref{assum:w_n_equal_v_n} hold. Then there exists a sequence of neural networks $\{f_n\}_{n=1}^\infty$ such that $J(f_n) \to 0$ and $f_n\to u$ strongly in $H^{0,1}(\Omega_T)$.
\end{theorem}

\subsection{Convergence in Unbounded Domain}
Define the $d$-dimensional ball centered at $0 \in \bR^d$ as $\Omega^k := \{x \in\bR^d: \sqrt{\sum_{i=1}^d x_i^2} < k\}$. $\partial \Omega^k$ is sufficiently smooth and satisfies Assumption \ref{assum:assumptions_on_Theorem_8.2}. Consider the following variational inequality with a variable bounded domain $\Omega_T^k := (0, T) \times \Omega^k$:
\begin{align}
    \label{eq:bounded_VI}
    \begin{cases}
        \min\{\cG[u^k](t,x), u^k(t,x)-g(t,x)\} = 0, & \text{for } (t,x) \in \Omega_T^k, \\
        u^k(t,x) = g(t,x), & \text{for } (t,x) \in \partial_p \Omega_T^k.
    \end{cases}
\end{align}
We require Assumption \ref{assum:assumptions_on_Theorem_8.2} hold for any $k\in\bN$. Then $u^k$ is unique and $u^k \in C(\overline{\Omega_T^k}) \cap W_p^{1,2}(\Omega_T^k)$ for any $1 < p < +\infty$ by Theorem 8.2 in Chapter 1 of \cite{friedmanVariationalPrinciplesFreeboundary1982}. Under Assumptions \ref{assum:assumptions_on_Theorem_8.2} and \ref{assum:conditions_on_u}, there exists a sequence of neural networks $\{f_n^k\}_{n=1}^\infty$ that converges to $u^k$ in $H^{0,1}(\Omega_T^k)$.

Consider the following variational inequality with the unbounded domain $\Omega_T^{\infty} := (0, T) \times \bR^d$:
\begin{align}
    \label{eq:unbounded_VI}
    \begin{cases}
        \min\{\cG[u^\infty](t,x), u^\infty(t,x)-g(t,x)\} = 0, & \text{for } (t,x) \in \Omega_T^{\infty}, \\
        u^\infty(0,x) = g(0,x), & \text{for } x \in \bR^d.
    \end{cases}
\end{align}
The solution to \eqref{eq:unbounded_VI} may not exist or be unique. Assumption \ref{assum:interior_estimates_for_u^k} below  is necessary for the maximum principle and interior estimates to construct $u^\infty$ from $\{u^k\}_{k=1}^\infty$ and could be satisfied by a stronger bound under certain growth conditions on $g$, see for example Lemma.A.2 in \cite{maGlobalClosedFormApproximation2019}.

\begin{assumption}
\label{assum:interior_estimates_for_u^k}
For all $ k > 0$, $|u^k(t,x)| \leq C(e^{\alpha x} + e^{-\gamma x})$ in $\Omega_T^k$, where  constants $C, \alpha, \gamma>0$ are independent of $k$.
\end{assumption}

\begin{theorem}
\label{thm:g_n_to_u_in_unbounded_domain}
Let Assumption \ref{assum:interior_estimates_for_u^k} holds. Suppose for any $k \in \bN$, Assumption \ref{assum:assumptions_on_Theorem_8.2} holds on $\Omega_T^k$. Then there exists a unique solution $u^\infty \in C(\overline{\Omega}_T^{\infty}) \cap W^{1,2}_{p, loc}(\Omega_T^{\infty})$ of \eqref{eq:unbounded_VI} for $1<p<+\infty$. 
Furthermore, suppose for any $k \in \bN$, on $\Omega_T^k$, Assumptions \ref{assum:conditions_on_u} to \ref{assum:w_n_equal_v_n} hold, then there exists a sequence $\{f_n^k\}_{n=1}^\infty$ such that $J(f_n^k)\to0$ and $f_n^k \to u^k$ strongly in $H^{0,1}(\Omega_T^k)$ as $n \to +\infty$.
Moreover, a single subsequence of $\{f_n^k\}_{n,k=1}^\infty$ strongly converges to $u^\infty$ in $H^{0,1}(M)$ on any compact set $M \in \Omega_T^\infty$.

\begin{proof}
    See \ref{app:proof_thm_5}.
\end{proof}
\end{theorem}

\begin{remark}
In contrast to the results for unbounded domain,  if the loss function tends to 0 for bounded domain, then the corresponding sequence of neural networks itself tends to the solution strongly in $H^{0,1}$, not a subsequence as in Theorem \ref{thm:g_n_to_u_in_unbounded_domain}. For many applications, even the domain is unbounded in theory, say $(0,\infty)$, we often restrict the domain to $(a,b)$ with $0<a<b<\infty$ in numerical computation, then we do not need to choose a subsequence.
\end{remark}

In the following, time is sometimes treated as a given parameter, and we would write $u(t;x)$ instead of $u(t,x)$. Corollary \ref{cor:g_n_to_u_a.a. time} gives the convergence of neural networks to the solution $u$ in spatial domain at almost any time.

\begin{corollary}
\label{cor:g_n_to_u_a.a. time}
Let the assumptions in Theorem \ref{thm:g_n_to_u_in_unbounded_domain} hold. Denote the sequence from Theorem \ref{thm:g_n_to_u_in_unbounded_domain} as $\{g_n\}_{n=1}^\infty$. Then there is a single subsequence of $\{g_n\}_{n=1}^\infty$, denoted as $\{v_n\}_{n=1}^\infty$, such that for almost any $t \in (0,T)$, $\{v_n(t;x)\}_{n=1}^\infty$ converges to $u^\infty(t;x)$ in $H^1(M)$ for any compact set $M \subset \bR^d$.

\begin{proof}
    See \ref{app:proof_cor_1}.
\end{proof}
\end{corollary}

\section{Application to Optimal Investment Stopping Problems} \label{sec:four}
In this section, we consider a special HJB type of nonlinear variational inequality arising from an optimal investment and stopping  problem, see \cite{maGlobalClosedFormApproximation2019} for detailed model setup and theoretical results on primal and dual problems and their relations. We next give a brief description.

Assume a complete Black-Scholes market model.  The wealth process of an investor satisfies the following SDE:
\[
dX_t = r X_t \, dt + \sigma \pi_t \left( \theta \, dt + dW_t \right),\ 0\leq t\leq T,
\]
where $r>0$ is the riskless interest rate, $\mu>0$  the stock growth rate, $\sigma >0$ the stock volatility, and $\theta = (\mu - r) / \sigma$ the market price of risk,  $(\pi_t)_{0 \leq t \leq T}$ is an $\mathcal{F}_t$-progressively measurable portfolio process satisfying $\mathbb{E}[\int_0^T |\pi_t|^2 dt] < \infty$.
The primal value function $V$ is defined by
\[
V(t, x) = \sup_{\tau, \pi} \mathbb{E} \left[ e^{-\beta(\tau - t)} U(X_\tau^{t, x, \pi} - K) \mid X_t = x \right], \ (t,x) \in Q_x:=(0,T) \times (K,+\infty),
\]
where  $\tau \in [0,T]$ is  an $\cF_t$-adapted stopping time, $\beta > 0$ a utility discount factor, $K$ a minimum wealth threshold value,  $U: [0,+\infty) \to [0,+\infty)$ is  $C^1$, increasing, strictly concave, and satisfies $U(0) = 0, \, \lim_{x\to\infty} U(x) = +\infty$, $U'(0) = +\infty$, $\lim_{x\to\infty} U'(x) = 0$, $U(x) < C(1+x^p)$ for $x \geq 0$ and some constant $C>0$, $0<p<1$, and $U(x) = -\infty$ for $x < 0$.

The primal value function $V$ satisfies the following nonlinear variational inequality:
\[        \min\{-\partial_t V + \frac{\theta^2}{2} \frac{(\partial_x V)^2}{\partial_{x}^2 V} - rx \partial_x V + \beta V, \, V - U(x-K)\} = 0, \ (t,x) \in Q_x,\]
with the terminal condition $V(T,x) = U(x-K)$ for $x \in (K,+\infty)$.
It is in general difficult to solve this nonlinear variational inequality. We use  the dual method described as follows.

 Define the dual process $(Y_t)_{0\leq t \leq T}$ by the SDE
\[
dY_t = (\beta - r) Y_t \, dt - \theta Y_t \, dW_t, \ 0\leq t\leq T,
\]
and the dual value function by
\[
\tilde{V}(t, y) := \sup_{t \leq \tau \leq T} \mathbb{E} \left[ e^{-\beta(\tau - t)} \tilde{U}_K(Y_\tau) \mid Y_t = y \right], \ (t,y) \in Q_y := (0,T) \times (0,+\infty),
\]
where $\tilde{U}_K(y) := \sup_{x>K}[U(x-K)-xy]$, $y>0$, is the dual utility function that is $C^1$, decreasing, strictly convex.  
The dual value function $\tilde{V}$ satisfies the following linear variational inequality:
\begin{equation} \label{eq:dual_VI}
           \min\{-\partial_t \tilde{V} - \frac{\theta^2}{2}y^2 \partial_y^2 \tilde{V} - (\beta-r) y \partial_y \tilde{V} + \beta \tilde{V}, \, \tilde{V} - \tilde{U}_K\} = 0, \ (t,y) \in Q_y
\end{equation}
with the terminal condition $\tilde{V}(T,y) = \tilde{U}_K(y)$ for $y \in (0,+\infty)$. Under some mild conditions on $\tilde{V}$, 
the  primal value function $V$ and the dual value function $\tilde{V}$ are related by
\begin{align} \label{eq:dual_transform}
V(t,x) = \inf_{y>0} \left\{ \tilde{V}(t,y) + xy \right\}, \ (t,x) \in Q_x.
\end{align}
We can also characterize the 
optimal stopping time $\tau^*$, optimal control $\pi^*$ and optimal wealth $X^*$ in terms of the dual process $Y$ and the dual value function $\tilde{V}$ and its derivatives, see, for example,  Theorems 1 and 3 in \cite{jangOptimalInvestmentHeterogeneous2024}.

 To further simplify the linear variational inequality (\ref{eq:dual_VI}), define 
\begin{align}
    \label{def:variable_transform}
z = \log y, \ \tau = \frac{\theta^2}{2}(T-t), \ v(\tau,z) = \tilde{V}(t,y).
\end{align}
Then $v$ satisfies a linear parabolic variational inequality:
\begin{equation}
    \label{eq:v_VI}
          \min\{\partial_\tau v - \partial_z^2 v + \kappa \partial_z v + \rho v, v(\tau,z)-\tilde{U}_K(e^z)\} = 0, \ (\tau,z) \in Q_z:=(0, \frac{\theta^2 T}{2}) \times \bR,
\end{equation}
with the initial condition $ v(0,z) = \tilde{U}_K(e^z)$ for $z \in \bR$, where  $\rho = 2\beta / \theta^2$ and $\kappa = (2r-2 \beta)/\theta^2+1$.

 The Sobolev embedding from $H^{1}(\Omega)$ to $C^{0,\frac{1}{2}}(\overline{\Omega})$ is applicable in this case. We can prove that if a sequence of neural networks converges to the transformed dual value function $v(\tau;z)$ in $H^{1}(M')$ for any compact set $M' \subset \bR$ at almost any time, then the corresponding approximations for primal value function also converges to $V(t;x)$ in $C^0(\overline{M})$ for any compact set $M \subset (K,+\infty)$ at almost any time and thus pointwise convergence holds in $(K,+\infty)$.

We first prove the convergence holds after the transformation from $v(\tau,z)$ to $\tilde{V}(t,y)$ because \eqref{def:variable_transform} is a homeomorphism \footnote{A homeomorphism between two topological spaces is a function that is (1) continuous, (2) bijective, and (3) has a continuous inverse. Being a homeomorphism implies that the function preserves topological properties such as open/closed sets, connectedness, and compactness.}.

\begin{proposition}
\label{prop:convergence_after_variable_transform}
For problem \eqref{eq:v_VI}, suppose Assumption \ref{assum:interior_estimates_for_u^k} holds, and for any $k\in\bN$, on $(0,\frac{\theta^2 T}{2}) \times (-k,k)$, Assumption \ref{assum:assumptions_on_Theorem_8.2} to \ref{assum:w_n_equal_v_n} holds, then from Theorem \ref{thm:g_n_to_u_in_unbounded_domain}, there is a single sequence of neural networks $\{g_n\}_{n=1}^\infty$ that converges to $v$ strongly in $H^{0,1}(M')$ for any compact set $M' \subset (0,\frac{\theta^2 T}{2}) \times \bR$. Define $h_n(t, y) = g_n(\tau, z)$ according to \eqref{def:variable_transform}. Then $\{h_n\}_{n=1}^\infty$ converges to the dual value function $\tilde{V}(t,y)$ strongly in $H^{0,1}(M)$ for any compact set $M \subset Q_y$.
\begin{proof}
See \ref{app:proof_prop_2}.
\end{proof}
\end{proposition}

Similar to Corollary \ref{cor:g_n_to_u_a.a. time}, the following corollary gives the convergence in spatial domain at almost any time.
\begin{corollary}
\label{cor:convergence_after_variable_transform_a.a. time}
Let the assumptions in Proposition \ref{prop:convergence_after_variable_transform} hold. Then there is a single subsequence of $\{h_n(t,y)\}_{n=1}^\infty$, denoted as $\{w_n(t,y)\}_{n=1}^\infty$, such that for almost any $t \in (0,T)$, $w_n(t;y)$ converges to the solution $\tilde{V}(t;y)$ in $H^1(B)$ for any compact set $B \subset (0, +\infty)$. Moreover,  $\{w_n(t;y)\}_{n=1}^\infty$ converges to $\tilde{V}(t;y)$ in $C^{0, \frac{1}{2}}(\overline{B})$.
\begin{proof}
    The proof of the first part is similar to Corollary \ref{cor:g_n_to_u_a.a. time}. The second part follows from the Sobolev embedding theorem due to  $y$ being a one-dimensional variable.
\end{proof}
\end{corollary}

Next, we prove the convergence to the primal value function by \eqref{eq:dual_transform}. From Corollary 2.4 in \cite{maGlobalClosedFormApproximation2019}, 
$\tilde{V}(t,y) \in C([0,T] \times (0, +\infty)) \cap W^{1,2}_{p,loc}([0,T) \times (0, +\infty))
$ 
for $1 < p < +\infty$. $\tilde{V}$ is decreasing in $t$ and strictly convex in $y$. For $(t,y) \in [0,T] \times (0, +\infty)$, 
\begin{align}
    \label{eq:limit_of_dual}
\lim_{y \to 0} \partial_y \tilde{V}(t,y) = -\infty, \quad \lim_{y \to +\infty} \partial_y \tilde{V}(t,y) \geq -K.
\end{align}
Therefore, there is a unique minimizer in \eqref{eq:dual_transform}, denoted as $y_x^t$, such that 
\[
y_x^t \in (0,+\infty), \quad \partial_y \tilde{V}(t,y_x^t)=-x, \quad V(t,x) = \tilde{V}(t,y_x^t) + xy_x^t.
\]
We prepick a compact domain for $x$, denoted as $M_x \subset (K, +\infty)$. Since the two limits of $\partial_y \tilde{V}$ in \eqref{eq:limit_of_dual} hold for a compact region of time, there exists a time-uniform compact range $M_y(M_x) \subset (0, +\infty)$ including all $y_x^t$ across $[0,T] \times M_x$. From Corollary \ref{cor:convergence_after_variable_transform_a.a. time}, we have a single sequence of neural networks $\{w_n(t,y)\}_{n=1}^\infty$. Define 
\begin{align}
\label{def:V_n}
V_n(t,x) := \inf_{y \in M_y(M_x)}\{w_n(t,y)+xy\}.
\end{align}
Since the neural network function $w_n$ is continuous on any compact set within $Q_y$, there is a minimizer $y_{x,n}^t \in M_y(M_x)$.

\begin{theorem}
\label{thm:uniform_convergence_to_primal_value}
Let the assumptions in Proposition \ref{prop:convergence_after_variable_transform} hold. Then for almost any $t \in [0,T]$, $\{V_{n}(t;x)\}_{n=1}^\infty$ converges to $V(t;x)$ uniformly on $M_x$ for any compact set $M_x \subset (K, +\infty)$. Moreover, $\{V_{n}(t,x)\}_{n=1}^\infty$ converges to $V(t,x)$ in $L^2(0,T; L^\infty(M_x))$.

\begin{proof}
    See \ref{app:proof_thm_7}.
\end{proof}
\end{theorem}

\section{Numerical Examples} \label{sec:five}
\subsection{Mixed Optimal Stopping and Control Problem}
Following Section \ref{sec:four}, we design the neural network to compute the primal value function and compare it with different methods. Two types of utility functions are considered:
power utility $U(x) = x^\gamma/\gamma$ for $0<\gamma<1$ and  non-HARA utility $U(x) = H^{-3}(x)/3 + H^{-1}(x) + xH(x)$, where $H(x) = \sqrt{2}(-1+\sqrt{1+4x})^{-1/2}$. 
It can be verified  that both dual utility functions satisfy  Assumption \ref{assum:assumptions_on_Theorem_8.2} on any compact set in $Q_z$. Assumption \ref{assum:interior_estimates_for_u^k} is satisfied as shown in Lemma A.2 in \cite{maGlobalClosedFormApproximation2019}.

In Table \ref{tab:1}, NN Bisection means we choose $y^*$ that minimizes $w_n(t,y) + x y$ from the bisection method towards $\partial_y(w_n(t,y) + x y)$. NN Grid means we choose $y^*$ by evaluating the value on a dense grid of 200 points of $y$ and choose the minimizer. The first method to compare is BTM (Binomial Tree Method), which solves the optimal stopping problem of the dual value function, and then get the primal value with bisection method. The other one is GCA (Global Closed-Form Approximation) from \cite{maGlobalClosedFormApproximation2019}, which approximates the free-boundary to compute the dual value, and attains the primal value with bisection method.

\begin{example}
\label{exa:1}
For power and non-HARA utility, we compare the numerical results among NN Bisection, NN Grid, GCA and BTM. The parameters used are $\mu = 0.1$, $\beta = 0.1$, $r = 0.05$, $\sigma = 0.3$, $K = 1$, $\gamma = 0.5$, $T = 1$, $t = 0$, initial wealth $x_0 \in [1.1, 1.2, 1.3, 1.4, 1.5, 1.6, 1.7, 1.8, 1.9, 2.0]$, number of time steps for BTM $N = 2000$.
\end{example}

\begin{figure}[H]
    \centering
    \begin{subfigure}[b]{0.48\textwidth}
        \centering
        \includegraphics[width=\textwidth]{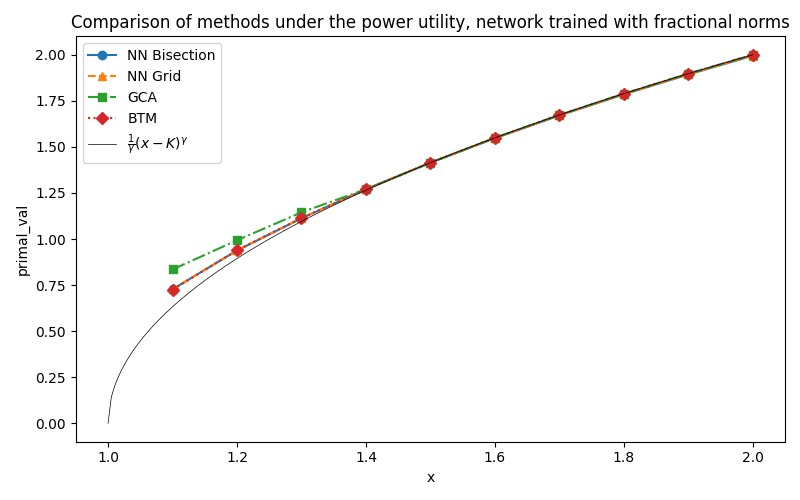}
        \caption{Power utility}
        \label{fig49}
    \end{subfigure}
    \hfill
    \begin{subfigure}[b]{0.48\textwidth}
        \centering
        \includegraphics[width=\textwidth]{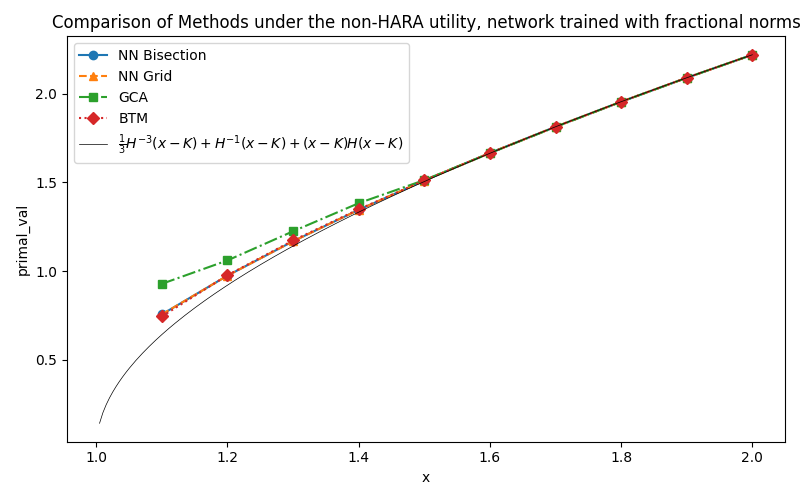}
        \caption{non-HARA utility}
        \label{fig17}
    \end{subfigure}
    \caption{Comparison of methods under different utilities}
    \label{fig:1}
\end{figure}

Figure \ref{fig:1} gives the comparisons in Example \ref{exa:1}. The black line indicates the immediate exercise value $U(x-K)$, and the gaps between it and other lines indicate the approximated time values. Table \ref{tab:1} illustrates the comparisons with numbers and the benchmark is BTM.

\begin{table}[H]
    \centering
    \caption{Comparison of methods with BTM as benchmark under different utilities}
    \label{tab:1}
    \begin{tabular}{ll *{4}{c}}
        \toprule
        Utility & Method & {\makecell{Mean Abs. \\ Rel. Diff.$^*$ (\%)}} & {\makecell{Std Rel. \\ Diff.$^\dagger$ (\%)}} & {\makecell{Training \\ Time (s)}} & {\makecell{Evaluation Time \\ /Point (ms)}} \\
        \midrule
        \multirow{3}{*}{Power} 
        & NN Bisection & 0.138755 & 0.140288 & 836.21 & 22\\
        & NN Grid & 0.138503 & 0.140122 & 836.21 & 21 \\
        & GCA & 2.379369 & 4.855817 & - & 19931 \\
        \midrule
        \multirow{3}{*}{non-HARA} 
        & NN Bisection & 0.249797 & 0.535413 & 845.49 & 19 \\
        & NN Grid & 0.249643 & 0.535435 & 845.49 & 18 \\
        & GCA & 4.051016 & 7.797414 & - & 32551 \\
        \bottomrule
    \end{tabular}
    \begin{tablenotes}
        \item{*}{Mean of the absolute value of relative differences across 10 points.}
        \item{$\dagger$}{Standard deviation of the relative differences across 10 points.}
    \end{tablenotes}
\end{table}

Table \ref{tab:1} shows the neural network method is much more accurate than GCA. The training time of neural networks is longer than the total evaluation time of GCA at 10 pairs of points, but still acceptable. Most importantly, once we have trained the network, it is instant to get $V(t,x)$ and its derivatives for different pairs of $(t,x)$, so the computation time remains the same even if we evaluate it on more points whereas the whole processes of both GCA and BTM would have to be repeated for different pairs of $(t,x)$. We also see that the NN Bisection and NN Grid methods are close to each other, which indicates that the bisection method works well, consistent with our theoretical conclusion that the first order derivative of the neural network converges.

\subsection{High-Dimensional American Option Pricing}
We test our neural network method on the American put option pricing problem in high dimensions. Assume there are $d$ risky assets with price processes $S_t^i$, $i=1,2,\ldots,d$, which follow the risk-neutral dynamics:
\begin{align*}
    S_t^i=s_0^i \exp\left\{\left(r-\delta_i-\frac{1}{2}\sigma_i^2\right)t + \sigma_i W_t^i\right\}, \quad i=1,2,\ldots,d,
\end{align*}
where $r$ is the constant riskless interest rate, $\delta_i$ the dividend yield, $\sigma_i$ the volatility, $s_0^i$ the initial value, and $(W^1,\dots,W^d)$ a $d$-dimensional Brownian motion with a constant correlation matrix $(\rho_{ij})_{i,j}$. Consider the payoff that depends on the product of the stock prices, given by
\begin{align*}
    \phi(\mathbf{s}) = \left(K-\Pi_{i=1}^d s^i\right)^{+}, \quad \mathbf{s} = (s^1, s^2, \ldots, s^d) \in (0,+\infty)^d,
\end{align*}
with strike $K$ and maturity $T$.
The value function of the American put option is given by
\begin{align*}
    V(t, \mathbf{s}) = \sup_{\tau\in\cT_{t,T}} \mathbb{E} \left[ e^{-r(\tau-t)} \phi(\mathbf{S_\tau}) \mid S_t^i = s^i, i=1,2,\ldots,d \right],
\end{align*}
where $\cT_{t,T}$ is the set of stopping times in $[t,T]$. The value function satisfies the following variational inequality:
\begin{align} \label{eq:VI_for_max_call}
           \min\{-\partial_t V -\cL_s V + r V, V - \phi\} = 0, \ (t, \mathbf{s}) \in [0,T] \times (0,+\infty)^d,
\end{align}
with the terminal condition $V(T, \mathbf{s}) = \phi(\mathbf{s})$ for $\mathbf{s} \in (0,+\infty)^d$,
where $\cL$ is the infinitesimal generator of the $d$-dimensional Brownian motion, given by
\begin{align*}
    \cL_s V = \frac{1}{2} \sum_{i,j=1}^d \sigma_i \sigma_j \rho_{ij} s^i s^j \partial_{s^is^j}V + \sum_{i=1}^d (r - \delta_i) s^i \partial_{s^i}V.
\end{align*}
There is a unique solution to \eqref{eq:VI_for_max_call}, see \cite{broadieValuationAmericanOptions1997} and \cite{jailletVariationalInequalitiesPricing1990a} for details. We shall turn to the log-price version for easier numerical implementations. Let $x^i=\log s^i$. The payoff function becomes
\begin{align*}
    g(\mathbf{x}) = \left(K-\exp{\left(\sum_{i=1}^d x^i\right)}\right)^{+}, \quad \mathbf{x} = (x^1, x^2, \ldots, x^d) \in \bR^d,
\end{align*}
and the value function becomes
\begin{align*}
    u(t, \mathbf{x}) = \sup_{\tau\in\cT_{t,T}} \mathbb{E} \left[ e^{-r(\tau-t)} g(\mathbf{X_\tau}) \mid X_t^i = x^i, i=1,2,\ldots,d \right].
\end{align*}
The value function $u(t,\mathbf{x})=V(t,e^\mathbf{x})$ where $\mathbf{x} \in \bR^d$ and $e^\mathbf{x}$ is defined componentwisely as $(e^{x^1}, e^{x^2}, \ldots, e^{x^d})$.
$u(t,\mathbf{x})$ satisfies the following variational inequality:
\begin{align} \label{eq:VI_for_max_call_log_price}
          \min\{-\partial_t u -\cL_x u + r u, u - g\} = 0, \ (t, \mathbf{x}) \in [0,T] \times \bR^d
\end{align}
with the terminal condition $u(T, \mathbf{x}) = g(\mathbf{x})$ for $\mathbf{x} \in \bR^d$,
where
\[
\cL_x u = \frac{1}{2} \sum_{i,j=1}^d \sigma_i\sigma_j \rho_{ij} \partial_{x^i x^j} u + \sum_{i=1}^d (r-\delta_i-\frac{1}{2}\sigma_i^2)\partial_{x^i}u.
\]
We test our method using the same parameters as in Section 5.3 in \cite{hureDeepBackwardSchemes2020}.
\begin{example}
\label{exa:2}
$K=1$, $T=1$, $r=0.05$, $\delta_i=\delta=0$, $s_0^i=s_0=1$, $\sigma_i=\sigma=0.2$ for $i=1,2,\dots,d$. $\rho_{ij}=\rho=0$ for $i\neq j$. The dimensions are $d=1,5,10,20$.
\end{example}
The results are shown in Table \ref{tab:2}.
\begin{table}[htbp]
    \centering
    \begin{threeparttable}
    \caption{Comparison of methods for high-dimensional American put option pricing}
    \label{tab:2}
    \begin{tabular}{l *{4}{c}}
        \toprule
        Dimension $d$ & {1} & {5} & {10} & {20} \\
        \midrule
        NN & 0.061458$\pm$0.00015 & 0.10693$\pm$0.00028 & 0.12970$\pm$0.00027 & 0.15033$\pm$0.00062 \\
        RDBDP & 0.061382$\pm$0.00019 & 0.10765$\pm$0.00016 & 0.12992$\pm$0.00016 & 0.15050$\pm$0.00010 \\
        Reference & 0.060903 & 0.10738 & 0.12996 & 0.1510 \\
        \bottomrule
    \end{tabular}
    \begin{tablenotes}
        \item{1.} The results are displayed as $\mu\pm\sigma$ across random seeds.
        \item{2.} NN: Our method. Trained with 5 consecutive random seeds.
        \item{3.} RDBDP: Reflected Deep Backward Dynamic Programming, see \cite{hureDeepBackwardSchemes2020}. Here are the results of RDBDP with 80 time steps and 40 random seeds.
        \item{4.} Reference: See Table 11 in \cite{hureDeepBackwardSchemes2020}.
    \end{tablenotes}
    \end{threeparttable}
\end{table}

\section{Proofs}\label{app:proofs}
To make the proofs simpler, we denote $K$ as a generic constant independent of $n$, $\epsilon$, $t$ and $x$, and it may be changed in different parts of the paper.
\subsection{Proof of Theorem \ref{thm:loss_function_convergence}} \label{app:proof_thm_1}
\begin{proof}
    To make the proof of Theorem \ref{thm:loss_function_convergence} simpler, we use the following bound, which follows from Assumption \ref{assum:assumptions_on_Theorem_8.2}: There exists $K>0$ such that for any $(t,x) \in \overline{\Omega}_T$,
    \begin{align*}
    |\cG[f](t,x)| \leq & K\left(|f(t,x)| + |\partial_t f(t,x)| + \sum_{i=1}^d|\partial_{x_i} f(t,x)| + \sum_{i,j=1}^d|\partial_{x_i x_j} f(t,x)|\right) \\
    \leq & K \left(\sup_{(t,x) \in \overline{\Omega}_T}|\partial_t f(t,x)| + \max_{|\alpha| \leq 2} \sup_{(t,x) \in \overline{\Omega}_T} |D_x^{\alpha} f(t,x)|\right).
    \end{align*}
    We choose the network $f \in \fC(\psi)$ that satisfies \eqref{eq:2-dense_property}. Consider the following term first:
    \begin{align*}
    & \|\min\{\cG[f], f-g\}\|^2_{L^2(\Omega_T), \mu_1} \\
    = & \|\min\{\cG[f], f-g\} - \min\{\cG[u], u-g\}\|^2_{L^2(\Omega_T), \mu_1} \\
    = & \|\cG[f]-\cG[u]\|^2_{L^2(S_1), \mu_1} + \|f - u\|^2_{L^2(S_2), \mu_1} + \|\cG[f]\|^2_{L^2(S_3), \mu_1} + \|f-g\|^2_{L^2(S_4), \mu_1},
    \end{align*}
    where
    \[S_1 = \{(t,x) \in \Omega_T \, | \, \cG[f](t,x;\theta) < f(t,x;\theta) - g(t,x), \, 0=\cG[u](t,x) < u(t,x) - g(t,x)\},\]
    \[S_2 = \{(t,x) \in \Omega_T \, | \, \cG[f](t,x;\theta) \geq f(t,x;\theta) - g(t,x), \, \cG[u](t,x) \geq u(t,x) - g(t,x) = 0\},\]
    \[S_3 = \{(t,x) \in \Omega_T \, | \, \cG[f](t,x;\theta) < f(t,x;\theta) - g(t,x), \, \cG[u](t,x) \geq u(t,x) - g(t,x) = 0\},\]
    \[S_4 = \{(t,x) \in \Omega_T \, | \, \cG[f](t,x;\theta) \geq f(t,x;\theta) - g(t,x), 0=\cG[u](t,x) < u(t,x) - g(t,x)\}.\]

    For the original equation \eqref{eq:original_VI}, $S_1 \cup S_4 = \cC$ is the continuation region and $S_2 \cup S_3 = \cS$ is the stopping region. In $S_1$,
    \begin{equation}
        \label{eq:bound_S_1}
    \begin{aligned}
    & \|\cG[f] - \cG[u]\|^2_{L^2(S_1), \mu_1} \\
    = & \int_{S_1} \cG[f - u]^2(t,x;\theta) \, d\mu_1(t,x) \\
    \leq & K\left(\sup_{(t,x) \in \overline{\Omega}_T}|\partial_t f(t,x;\theta) - \partial_t u(t,x)|^2 + \max_{|\alpha|\leq 2} \sup_{(t,x) \in \overline{\Omega}_T}|D_x^\alpha f(t,x;\theta) - D_x^\alpha u(t,x)|^2 \right) \\
    < & K\epsilon^2,
    \end{aligned}
    \end{equation}
    where we used the bound from \eqref{eq:2-dense_property}. With the same reason, in $S_2$, \[\|f - u\|^2_{L^2(S_2), \mu_1} < \epsilon^2.\]
    In $S_3$, from its definition and \eqref{eq:2-dense_property},\[ \cG[f](t,x;\theta) < f(t,x;\theta) - g(t,x) = f(t,x;\theta) - u(t,x) < \epsilon.\]
    On the other side, 
    \begin{align*}
    & |\cG[f](t,x;\theta) - \cG[u](t,x)| \\
    \leq & K\left(|\partial_t f(t,x;\theta) - \partial_t u(t,x)| + \max_{|\alpha|\leq2} |D_x^\alpha f(t,x;\theta)-D_x^\alpha u(t,x)|\right) \\
    < & K\epsilon.
    \end{align*}
    From the definition of $S_3$, $\cG[u](t,x) \geq 0$, so $\cG[f](t,x;\theta) > \cG[u](t,x) - K\epsilon \geq -K\epsilon$. Combining the two inequalities, we have $-K\epsilon \leq \cG[f](t,x;\theta) < \epsilon$, so \[\|\cG[f]\|^2_{L^2(S_3), \mu_1} \leq K^2\epsilon^2.\]
    In $S_4$, $u(t,x) - g(t,x) > 0$. From \eqref{eq:2-dense_property}, $f(t,x;\theta) - u(t,x) > -\epsilon$. Hence, \[f(t,x;\theta) - g(t,x) > u(t,x) - g(t,x) - \epsilon > - \epsilon.\] Also, in $S_4$, $\cG[u](t,x) = 0$, so it is in the continuation region. We can repeat the computations in \eqref{eq:bound_S_1} and get 
    \[
    \|\cG[f]\|^2_{L^2(S_4), \mu_1} < K \epsilon^2.
    \]
    In $S_4$, $\cG[f](t,x;\theta) \geq f(t,x;\theta) - g(t,x)$, so we know $\cG[f](t,x;\theta) + \epsilon \geq f(t,x;\theta) - g(t,x) + \epsilon > 0$. Hence,
    \begin{align*}
    & \|f - g\|^2_{L^2(S_4), \mu_1} \\
    \leq & 2 \|f - g + \epsilon\|^2_{L^2(S_4), \mu_1} + 2 \epsilon^2 \\
    \leq & 2 \|\cG[f] + \epsilon\|^2_{L^2(S_4), \mu_1} + 2 \epsilon^2 \\
    \leq & 4 \|\cG[f]\|^2_{L^2(S_4), \mu_1} + 6\epsilon^2 \\
    < & K \epsilon^2.
    \end{align*}

    For the second term in \eqref{def:J_f}, since $g(t,x) = u(t,x)$ on $\partial_p\Omega_T$, we have
    \begin{align*}
    \|e\|^2_{H^{1}(\Omega), \mu_2} &= \|f(0,\cdot \,;\theta)-u(0,\cdot)\|^2_{H^{1}(\Omega), \mu_2}\\
    &\leq K\epsilon^2.
    \end{align*}

    Similarly, the last two terms in \eqref{def:J_f} can also be bounded by $K\epsilon^2$ since the universal approximation theorem guarantees a higher order of uniform convergence.

    Combining the bounds above together, and after rescaling $\epsilon$, we have $J(f) < \epsilon$. We emphasize that the existence of such $f$ is directly from the universal approximation Theorem 3 in \cite{hornikApproximationCapabilitiesMultilayer1991} with respect to the solution to \eqref{eq:original_VI}, so there is no mixture of two different existence results.
    
\end{proof}

\subsection{Proof of Theorem \ref{thm:hat_f_n_converges_to_u}} \label{app:proof_thm_2}
\begin{proof}
    As we said before, $u$ is unique and $u \in C(\overline{\Omega}_T) \cap W_p^{1,2}(\Omega_T)$ for any $1 < p < +\infty$ by Theorem 1.8.2 in \cite{friedmanVariationalPrinciplesFreeboundary1982}. From the same idea in Problem 1 in Section 1.3 of \cite{friedmanVariationalPrinciplesFreeboundary1982} in the elliptic case, under Assumption \ref{assum:assumptions_on_Theorem_8.2}, since $v_n \to 0$ in $H^{1,2}(\Omega_T)$, $\hat{f}_n$ exists and is unique and $\hat{f}_n \in H^{1,2}(\Omega_T)$ for any $n$. Moreover, by the uniform $H^{1,2}(\Omega_T)$ bound of $\{v_n\}$, the $H^{1,2}(\Omega_T)$ bound of $\{\hat{f}_n\}$ is uniform. For details of the proof in the elliptic case, see Lemma 4.5.1 and Theorem 4.5.2 in \cite{kinderlehrerIntroductionVariationalInequalities1980}. For the parabolic case, see Proposition 2.1 and 2.2 in \cite{adamsOptimalControlObstacle2002}. 
    
    Since $H^{1,2}(\Omega_T)$ is reflexive (all Hilbert spaces are reflexive), there exists a weak limit $\hat{f}$ such that a subsequence of $\hat{f}_n$ converges weakly to $\hat{f}$ in $H^{1,2}(\Omega_T)$. By Rellich-Kondrachov theorem (Theorem 9.16 in \cite{brezisFunctionalAnalysisSobolev2011} and Proposition 4.4 in \cite{taylorPartialDifferentialEquations2023}), there exist two compact embeddings \footnote{Let \(X\) and \(Y\) be Banach spaces.  \(X\) is said to be {\it continuously embedded} in \(Y\)
    there exists a constant \(C>0\) such that  $    \|x\|_Y \leq C\,\|x\|_X$ for all  $x \in X$.
    The embedding \(X \hookrightarrow Y\) is said to be \emph{compact} if the inclusion operator
  $    I : X \to Y$ with $I(x)=x$ 
    is a compact operator, that is,  for every bounded sequence \(\{x_n\}\) in \(X\), there exists a subsequence \(\{x_{n_k}\}\) and an element \(y \in Y\) such that $    \lim_{k \to \infty} \|x_{n_k} - y\|_Y = 0$;
equivalently, the image of any bounded set in \(X\) is relatively compact in \(Y\). We use the notation \(X \subset \subset Y\) to indicate that the embedding is compact.}
    : 
    \[W^{2,2}(\Omega) \subset\subset W^{1,2}(\Omega) \subset\subset L^2(\Omega).\]
    We already know $\hat{f}_n \in H^{1,2}(\Omega_T)$, which is equivalent to $\hat{f}_n \in L^2(0,T; W^{2,2}(\Omega))$ and $\partial_t \hat{f}_n \in L^2(\Omega_T)$. Therefore, by the Aubin-Lions Theorem (see, e.g., Theorem II.5.16 in \cite{boyerMathematicalToolsStudy2013}), $\hat{f}_n \to \hat{f}$ strongly in $L^2(0, T;W^{1,2}(\Omega))$, i.e., $H^{0,1}(\Omega_T)$, up to subsequences. Therefore, 
    \[
    \hat{f}_n \to \hat{f} \text{ strongly in } L^2(\Omega_T), \quad D_x \hat{f}_n \to D_x \hat{f} \text{ strongly in } L^2(\Omega_T) \text{ up to subsequences}.
    \]
    
    Next, we prove $\hat{f} = u$ a.e. in $\Omega_T$. We firstly formulate the weak form of \eqref{eq:original_VI} for $u$. Define the admissible set of \eqref{eq:original_VI} to be \[K := \{\phi \in H^1(\Omega_T), \phi = g \text{ on } \partial_p\Omega_T, \phi \geq g \text{ a.e.}\},\] 
    where $\phi=g$ on $\partial_p\Omega_T$ is taken in the trace sense, e.g. see Chapter 5.5 in \cite{evansPartialDifferentialEquations2010}. The weak form of \eqref{eq:original_VI} is the following: $u$ satisfies
    \begin{equation}
        \label{eq:weak_formulation}
    u \in K, \quad \int_\Omega \partial_tu \, (\phi-u) \, dx+ a(t;u,\phi-u) \geq 0, \quad \text{ for a.a. } t \in (0,T), \quad \forall \phi \in K,
    \end{equation}
    see Chapter 1.8 in \cite{friedmanVariationalPrinciplesFreeboundary1982}. From Theorem 1.8.2 in \cite{friedmanVariationalPrinciplesFreeboundary1982}, the function that satisfies \eqref{eq:weak_formulation} is unique. Hence, we only need to prove $\hat{f}$ satisfies \eqref{eq:weak_formulation}. It is proved by Lemma \ref{lem:step_1} and Lemma \ref{lem:step_2} in the following.
    
    \begin{lemma}
        \label{lem:step_1}
        Denote the strong limit of $\hat{f}_n$ in $H^{0,1}(\Omega_T)$ as $\hat{f}$. Then $\hat{f} \in K$.
    \begin{proof}
    Rewrite \eqref{eq:auxilliary_VI} as the following:
    \begin{equation}
        \label{eq:rewrite}
        \begin{cases}
            \hat{f}_n(t,x) \geq g(t,x) + v_n(t,x), & \text{for } (t,x) \in \Omega_T, \\
            \cG[\hat{f}_n](t,x) \geq v_n(t,x), & \text{for } (t,x) \in \Omega_T, \\
            (\cG[\hat{f}_n](t,x) - v_n(t,x)) (\hat{f}_n(t,x)-g(t,x)-v_n(t,x)) = 0, & \text{for } (t,x) \in \Omega_T, \\
            \hat{f}_n(t,x) = g(t,x) + v_n(t,x), & \text{for } (t,x) \in \partial_p \Omega_T.
        \end{cases}
    \end{equation}
    
    Since $v_n\to0$ in $H^{1,2}(\Omega_T)$ and $v_n\to 0$ in $L^2(\partial_p\Omega_T)$, letting $n \to +\infty$, a subsequence of $v_n$ converges to zero a.e. in $\overline{\Omega}_T$. If we choose the subsequence that $\hat{f}_n \to \hat{f}$ in $H^{0,1}(\Omega_T)$ and $v_n \to 0$ a.e. in $\overline{\Omega}_T$, then $\hat{f} \geq g$ on $\Omega_T$ and $\hat{f} = g$ on $\partial_p\Omega_T$ a.e. by \eqref{eq:rewrite}.
    
    Moreover, $\hat{f}\in H^1(\Omega_T)$ from it is in $H^{1,2}(\Omega_T)$. Hence, $\hat{f} \in K$ up to a modification in a zero-measure set, which does not affect the results since $K$ is defined in a weak sense.
    \end{proof}
    \end{lemma}

    Next, we prove a short lemma for the Leibniz integral rule in Sobolev space.
    \begin{lemma}
    \label{lem:exchange_of_differentiation_and_integration}
    If $f\in W^{1,1}(0,T;L^1(\Omega))$, then for almost any $t\in(0,T)$,
    \begin{align*}
        \frac{d}{dt}\int_\Omega f(t,x)\,dx = \int_\Omega \partial_t f(t,x) \, dx.
    \end{align*}
    \begin{proof}
        Define the bounded linear functional $\Lambda:L^{1}(\Omega)\to\mathbb{R}$ by
        \[
        \Lambda(u)\;:=\;\int_{\Omega} u(x)\,dx
        \]
        Note that $\Lambda$ is continuous with $|\Lambda(u)|\le \|u\|_{L^1(\Omega)}$. Let $f\in W^{1,1}(0,T;L^{1}(\Omega))$. By definition of the Bochner--Sobolev space, there exists $g\in L^{1}(0,T;L^{1}(\Omega))$ such that the distributional time derivative $\partial_t f = g$, i.e.,
        \begin{equation}\label{eq:bochner-derivative}
        \int_0^T \int_\Omega f(t,x) \varphi'(t)\,dx\,dt = -\int_0^T\int_\Omega g(t,x) \varphi(t)\,dx\,dt, \quad \forall \varphi\in C_c^{\infty}(0,T).
        \end{equation}
        Define $F:(0,T)\to\bR$ by $F(t):=(\Lambda\circ f)(t)=\Lambda\bigl(f(t)\bigr)=\int_\Omega f(t,x)\,dx$. Using \eqref{eq:bochner-derivative} together with the linearity and continuity of $\Lambda$, we obtain
        \begin{align*}
        \int_{0}^{T} F(t)\,\varphi'(t)\,dt
        &=\int_{0}^{T}\Lambda\bigl(f(t)\bigr)\,\varphi'(t)\,dt
        =\int_{0}^{T}\!\!\int_{\Omega} f(t,x)\,\varphi'(t)\,dx\,dt\\
        &= -\int_{0}^{T}\!\!\int_{\Omega} g(t,x)\,\varphi(t)\,dx\,dt
        = -\int_{0}^{T}\Lambda\bigl(g(t)\bigr)\,\varphi(t)\,dt.
        \end{align*}
        Hence, in the sense of distributions on $(0,T)$,
        \[
        F'(t)=\Lambda\bigl(g(t)\bigr)=\int_\Omega g(t,x)\,dx, \quad \text{for almost any } t.
        \]
        Since $g\in L^{1}(0,T;L^{1}(\Omega))$ and $\Lambda$ is bounded, we have
        $\Lambda(g(\cdot))\in L^{1}(0,T)$. Therefore, $F\in W^{1,1}(0,T)$ and $F$ admits a weak time derivative $F'(t)=\int_{\Omega} g(t,x)\,dx$.
        It is equivalent to
        \[
        \frac{d}{dt}\int_{\Omega} f(t,x)\,dx
        =F'(t)
        =\int_{\Omega} g(t,x)\,dx
        =\int_{\Omega} \partial_t f(t,x)\,dx,
        \quad\text{for almost any } t\in(0,T).
        \]

    \end{proof}
    \end{lemma}
    
    \begin{lemma}
    \label{lem:step_2}
    Denote the strong limit of $\hat{f}_n$ in $H^{0,1}(\Omega_T)$ as $\hat{f}$. Then $\hat{f}$ satisfies
    \begin{align*}
    \int_\Omega \partial_t \hat{f} \, (\phi-\hat{f}) + a(t;\hat{f},\phi-\hat{f}) \geq 0, \quad \text{ for a.a. } t \in (0,T), \quad \forall \phi \in K.
    \end{align*}
    \begin{proof}
        Define the admissible set of \eqref{eq:auxilliary_VI} to be \[K_n := \{\phi \in H^1(\Omega_T), \phi = g + v_n \text{ on } \partial_p\Omega_T, \phi \geq g + v_n \text{ a.e.}\},\]
    where $\phi = g + v_n$ on $\partial_p\Omega_T$ is taken in the trace sense. The weak form of \eqref{eq:auxilliary_VI} is the following: $\hat{f}_n$ satisfies
    \begin{equation*}
    \hat{f}_n \in K_n, \int_\Omega \partial_t \hat{f}_n\,(\phi-\hat{f}_n) \, dx \, + \, a(t;\hat{f}_n, \phi-\hat{f}_n) \geq \int_\Omega v_n (\phi-\hat{f}_n)\,dx, \text{ for a.a. } t \in (0,T), \forall \phi \in K_n.
    \end{equation*}
    Following the definition in the previous proof, $\forall \phi \in K,$ define 
    \[
    \phi_n := \phi + v_n.
    \]
    Since $v_n \to 0$ in $H^{1,2}(\Omega_T)$, we know $\phi_n \in K_n$ and $\phi_n \to \phi$ in $H^1(\Omega_T)$ as $n \to +\infty$.

    Let $t_1, t_2 \in (0,T)$ be arbitrary and $t_1<t_2$. From integration by parts, the following holds:
    \begin{align*}
        & \int_{t_1}^{t_2}\int_\Omega\partial_t \hat{f}_n(t,x)\phi_n(t,x) \,dx\,dt \\
        =&\int_{\Omega} \hat{f}_n(t_2, x) \phi_n(t_2,x) \,dx 
        - \int_{\Omega} \hat{f}_n(t_1, x) \phi_n(t_1,x) \,dx-\int_{t_1}^{t_2}\int_\Omega \hat{f}_n(t,x) \partial_t\phi_n(t,x) \,dx\,dt,
    \end{align*}
    and
    \begin{align*}
        \int_{t_1}^{t_2}\int_\Omega\partial_t \hat{f}_n(t,x)\hat{f}_n(t,x) \,dx\,dt 
        &= \frac{1}{2}\int_{\Omega} \Bigl(\hat{f}_n^2(t_2, x) - \hat{f}_n^2(t_1, x)\Bigr) \,dx.
    \end{align*}
    
    In the above, Fubini theorem is used to exchange the order of integration. It is valid since $\hat{f}_n \in H^{1,2}(\Omega_T)$ and $\phi_n \in H^1(\Omega_T)$. We also exchanged the differentiation and integration for $\hat{f}_n^2(t,x)$ based on Lemma \ref{lem:exchange_of_differentiation_and_integration}. Similarly,
    \begin{align*}
        &\int_{t_1}^{t_2}\int_\Omega\partial_t \hat{f}(t,x)\phi(t,x) \,dx\,dt \\
        =&\int_{\Omega} \hat{f}(t_2, x) \phi(t_2,x) \,dx
        - \int_{\Omega} \hat{f}(t_1, x) \phi(t_1,x) \,dx-\int_{t_1}^{t_2}\int_\Omega \hat{f}(t,x) \partial_t\phi(t,x) \,dx\,dt,
    \end{align*}
    and
    \begin{align*}
        \int_{t_1}^{t_2}\int_\Omega\partial_t \hat{f}(t,x)\hat{f}(t,x) \,dx\,dt 
        &= \frac{1}{2}\int_{\Omega} \Bigl(\hat{f}^2(t_2, x) - \hat{f}^2(t_1, x)\Bigr) \,dx.
    \end{align*}
    Since $v_n\to 0$ in $H^{1,2}(\Omega_T)$, $\phi_n\to\phi$ in $H^1(\Omega_T)$ and $\hat{f}_n\to\hat{f}$ in $H^{0,1}(\Omega_T)$, the above equations give that for almost any $t_1,t_2 \in (0,T)$, up to a subsequence,
    \[\int_{t_1}^{t_2}\int_\Omega\partial_t\hat{f}(\phi-\hat{f})\,dx\,dt=\lim_{n\to +\infty}\int_{t_1}^{t_2}\int_\Omega(\partial_t\hat{f}_n-v_n)(\phi_n-\hat{f}_n)\,dx\,dt.\]
    From Assumption \ref{assum:assumptions_on_Theorem_8.2}, $|D_x a_{ij}|\leq C$, $a_{ij}, b_i, c$ are bounded in H\"older's norm, so we have
    \begin{align*}
    \int_{t_1}^{t_2}\int_\Omega a(t;\hat{f},\phi-\hat{f}) \, dt = \lim_{n\to +\infty}\int_{t_1}^{t_2}\int_\Omega a(t;\hat{f}_n,\phi_n-\hat{f}_n) \, dt.
    \end{align*}
    Hence,
    \begin{align*}
    & \int_{t_1}^{t_2}\left(\int_\Omega \partial_t \hat{f} \, (\phi-\hat{f}) \, dx+ a(t;\hat{f},\phi-\hat{f})\right)\,dt\\
    = & \lim_{n \to +\infty} \int_{t_1}^{t_2}\left(\int_\Omega (\partial_t \hat{f}_n-v_n)\,(\phi-\hat{f}_n) \, dx + a(t;\hat{f}_n, \phi-\hat{f}_n)\right) \geq 0,
    \end{align*}
    for almost any $t_1,t_2 \in (0,T)$.
    Considering the term as an increment, the following Lebesgue integral is nondecreasing:
    \[
    F(\tau) := \int_0^\tau \left(\int_\Omega \partial_t \hat{f} \, (\phi-\hat{f}) \, dx+ a(t;\hat{f},\phi-\hat{f})\right)\,d\tau.
    \]
    From the absolute continuity of Lebesgue integral and the nondecreasing property, its derivative 
    \[
    \int_\Omega \partial_t \hat{f} \, (\phi-\hat{f}) \, dx+ a(t;\hat{f},\phi-\hat{f}) \geq 0
    \] 
    for almost any $t \in (0, T)$.
    \end{proof}
    \end{lemma}

    Combining Lemma \ref{lem:step_1} and Lemma \ref{lem:step_2}, $\hat{f}$ is the solution to \eqref{eq:weak_formulation} and $\hat{f}=u$. Since for all subsequences of $\hat{f}_n$, the above proof holds, so there is always a sub-subsequence of $\hat{f}_n$ that converges to the same limit $u$ in $H^{0,1}(\Omega_T)$, which concludes $\hat{f}_n \to u$ in $H^{0,1}(\Omega_T)$.
\end{proof}

\subsection{Proof of Theorem \ref{thm:f_n_converges_to_hat_f_n}} \label{app:proof_thm_3}
\begin{proof}
    For the penalty method of $f_{n}$, we first write its possible form with some penalty function to be specified later. We allow the penalty function $\beta_{n,\epsilon}^t$ to depend on time $t$ and different for each $n$. The penalized form of \eqref{eq:VI_f_n} is
    \begin{align} \label{eq:penalized_f}
    \begin{cases}
    \cG[f_{n, \epsilon}] + \beta_{n,\epsilon}^t(f_{n, \epsilon} - g - w_n) = w_n, & \text{for } (t,x) \in \Omega_T, \\
    f_{n,\epsilon} = g + v_n, & \text{for } (t,x) \in \partial_p\Omega_T.
    \end{cases}
    \end{align}
    Define $r_{n,\epsilon}:=f_{n,\epsilon}-g-v_n$, then we have
    \begin{align} \label{eq:penalized_r}
    \begin{cases}
    \cG[r_{n, \epsilon}] + \beta_{n,\epsilon}^t(r_{n, \epsilon} + v_n - w_n) = w_n- \cG[g+v_n], & \text{for } (t,x) \in \Omega_T, \\
    r_{n,\epsilon} = 0, & \text{for } (t,x) \in \partial_p\Omega_T.
    \end{cases}
    \end{align}
    Define $\varphi_n:=w_n-v_n$. Note that $\varphi_n(t,\cdot)$ has the same piecewise smooth partitions in $\Omega$ as $w_n$. We have
    \[
    \varphi_n(t,\cdot)|_{\Omega_{i,n}^t} \in H^2(\Omega_{i,n}^t), \quad 1\leq i \leq m_t^n, \quad \forall t \in (0,T).
    \]
    For each $k=1,\ldots,M_t^n$, suppose $\Gamma_{k,n}^t=\partial\Omega_{i,n}^t \cap \partial\Omega_{j,n}^t$ for some $1\leq i \neq j \leq m_t^n$, define
    \[
    q_{k,n}(t,x) = \nabla \varphi_n(t,x) |_{\overline{\Omega_{i,n}^t}} \cdot \eta_{i,n}^t(x) + \nabla \varphi_n(t,x) |_{\overline{\Omega_{j,n}^t}} \cdot \eta_{j,n}^t(x), \quad x \in \Gamma_{k,n}^t,
    \]
    where $\eta_{i,n}^t(x)$ and $\eta_{j,n}^t(x)$ are the outward normal unit vectors of the boundaries $\partial\Omega_{i,n}^t$ and $\partial\Omega_{j,n}^t$ at the point $x \in \Gamma_{k,n}^t$. Denote $s_n:=w_n-\cG[g+v_n]$. Define
    \[
    p_n \in L^2(\Omega_T): \quad p_n(t,\cdot)|_{\Omega_{i,n}^t} = (\cG[\varphi_n](t,\cdot) - s_n(t,\cdot))|_{\Omega_{i,n}^t}, \quad 1\leq i \leq m_t^n,
    \]
    \[
    \forall t \in (0,T), \quad q_n(t,\cdot) \in L^2(\Gamma_n^t): \quad q_n(t,\cdot)|_{\Gamma_{k,n}^t} = q_{k,n}(t,\cdot), \quad 1\leq k \leq M_t^n.
    \]
    Under Assumption \ref{assum:w_n}, the above functions are well-defined and are uniformly in $L^2$. Choose any two functions $G_n \geq p_n^+$ and $Q_n \geq q_n^+$ such that $\int_{\Gamma_n^t}Q_n(t,x) \, dx$ is measurable in $t$, and $G_n, Q_n$ are uniformly in $L^2$.

    Let $\theta_\epsilon (\xi) := -\xi^-/\epsilon$. The penalty function $\beta_{n,\epsilon}^t(\xi)$ is defined on $\bR$ for $0<\epsilon<1$ such that
    \begin{align}
        \label{eq:penalty_function}
    \langle \beta_{n,\epsilon}^t(\xi), \phi \rangle = \int_\Omega G_n(t,x) \theta_\epsilon(\xi)\phi(x) \, dx+ \int_{\Gamma_n^t} Q_n(t,x) \theta_\epsilon(\xi)\phi(x) \, dx, \quad \forall \phi \in H^1_0(\Omega),
    \end{align}
    where $\langle \cdot, \cdot \rangle$ is the duality pairing between $H^1_0(\Omega)$ and its dual space.

    With the above formulations, we adapt the following result from \cite{dautovPenaltyMethodsOneSided2021} to \eqref{eq:penalized_r}.
    \begin{lemma}[Theorems 3,4 and Example 2, \cite{dautovPenaltyMethodsOneSided2021}]
        \label{lem:dautov}
        Denote $r_n = f_n - g - v_n$. For any $n \in \bN$, applying the penalty function \eqref{eq:penalty_function} to \eqref{eq:penalized_r}, we have
        \[
         r_n-\epsilon \leq r_{n,\epsilon} \leq r_n, \quad \|r_{n,\epsilon} - r_n\|_{H^{0,1}(\Omega_T)} \leq C_n \epsilon^{1/2},
        \]
        where
        \[
        C_n = 2 \left( \int_0^T \int_\Omega G_n(t,x) \, dx \, dt + \int_0^T \int_{\Gamma_n^t} Q_n(t,x) \, dx \, dt \right).
        \]
    \begin{proof}
    The proof follows from the similar arguments as in \cite{dautovPenaltyMethodsOneSided2021}. We list the main changes here.
    \begin{itemize}
        \item The parabolic operator changes to a more complex form $\cG$. Since the bilinear form satisfies $a(t;u,u) \geq M \|u\|^2_{H^1(\Omega)}$, we can treat it similarly as in the original proof.
        \item The obstacle is now dependent on time, and piecewise smooth in $\Omega$ at any $t \in [0,T]$. This does not affect the overall structure of the proof. After putting $\partial_t \varphi_n$ into the definition of $p_n$ and allowing the set of test functions, the convex functionals, and the penalty operator to depend on time, we can follow the same steps as in the original proof.
        \item The original paper restricts the dimension $d \in \{2,3\}$, but this is only used in its Lemma 2 to ensure the local Lipschitz continuity of the functionals. Since we do not need this result in our setting, we can work in any spatial dimension.
    \end{itemize}
    \end{proof}
    \end{lemma}
From Lemma \ref{lem:dautov}, we have 
\begin{align} \label{eq:bound_f_n_epsilon}
    f_n-\epsilon \leq f_{n,\epsilon} \leq f_n, \quad \|f_{n,\epsilon} - f_n\|_{H^{0,1}(\Omega_T)} \leq C_n \epsilon^{1/2}.
\end{align}
Since $f_n \geq g+w_n$, we have $f_{n,\epsilon}-g-w_n \geq -\epsilon$ and $-1 \leq \theta_\epsilon(f_{n,\epsilon}-g-w_n) \leq 0$. Therefore, $\forall \phi \in H^1_0(\Omega)$,
\begin{equation}
\begin{aligned} \label{eq:penalty_estimate}
\left|\langle \beta_{n,\epsilon}^t(f_{n,\epsilon}-g-w_n), \phi \rangle \right| \leq & \int_\Omega \left|G_n(t,x) \phi(x)\right| \, dx+ \int_{\Gamma_n^t} \left|Q_n(t,x) \phi(x)\right| \, dx \\
\leq & K \|\phi\|_{L^2(\Omega)} + K \|\phi\|_{L^2(\Gamma_n^t)} \\
\leq & K \|\phi\|_{L^2(\Omega)} + K \|\phi\|_{L^2(\Omega)}^{1/2} \|\phi\|_{H^1(\Omega)}^{1/2},
\end{aligned}
\end{equation}
where the last inequality follows from Theorem 1.5.1.10 in \cite{grisvardEllipticProblemsNonsmooth1985} after taking the maximum over $\epsilon$.

For the penalty method of $\hat{f}_n$, we can repeat the same steps as for $f_n$. There are similar functions $\hat{G}_n$ and $\hat{Q}_n$ associated with $\hat{f}_n$. By choosing the maximum between $\hat{G}_n$ and $G_n$, $\hat{Q}_n$ and $Q_n$, we can ensure that the penalty function remains the same for both penalized problems. For simplicity, we preserve the notation for the penalty function as $\beta_{n,\epsilon}^t$, and the corresponding $G_n$ and $Q_n$ are defined accordingly.

The penalized form of \eqref{eq:auxilliary_VI} is
\begin{align} \label{eq:penalized_hat_f}
\begin{cases}
\cG[\hat{f}_{n, \epsilon}] + \beta_{n,\epsilon}^t(\hat{f}_{n, \epsilon} - g - v_n) = v_n, & \text{for } (t,x) \in \Omega_T, \\
\hat{f}_{n, \epsilon} = g + v_n, & \text{for } (t,x) \in \partial_p\Omega_T.
\end{cases}
\end{align}
Similarly to \eqref{eq:bound_f_n_epsilon} and \eqref{eq:penalty_estimate}, we have
\begin{align} \label{eq:bound_hat_f_n_epsilon}
\hat{f}_n - \epsilon \leq \hat{f}_{n,\epsilon} \leq \hat{f}_n, \quad \|\hat{f}_{n,\epsilon} - \hat{f}_n\|_{H^{0,1}(\Omega_T)} \leq C_n \epsilon^{1/2},
\end{align}
and
\[
\left|\langle \beta_{n,\epsilon}^t(\hat{f}_{n,\epsilon} - g - v_n), \phi \rangle \right| \leq K \|\phi\|_{L^2(\Omega)} + K \|\phi\|_{L^2(\Omega)}^{1/2} \|\phi\|_{H^1(\Omega)}^{1/2}.
\]

Define $h_n := f_n - \hat{f}_n$, $h_{n, \epsilon} := f_{n, \epsilon} - \hat{f}_{n, \epsilon}$, $z_{n, \epsilon} := f_{n, \epsilon} - g - w_n$, $\hat{z}_{n, \epsilon} := \hat{f}_{n, \epsilon} - g - v_n$, $e_n := w_n - v_n$. Taking the difference between \eqref{eq:penalized_f} and \eqref{eq:penalized_hat_f}, we have
\begin{align}
    \label{eq:PDE_for_h_{n,epsilon}}
    \begin{cases}
        \cG[h_{n, \epsilon}] + \beta_{n,\epsilon}^t(z_{n, \epsilon}) - \beta_{n,\epsilon}^t(\hat{z}_{n, \epsilon}) = e_n, & \text{for } (t,x) \in \Omega_T, \\
        h_{n, \epsilon} = 0, & \text{for } (t,x) \in \partial_p\Omega_T.
    \end{cases}
\end{align}

\begin{lemma}
    \label{lem:H1_convergence_of_h_{n,epsilon}}
    Let Assumptions \ref{assum:assumptions_on_Theorem_8.2}, \ref{assum:w_n} and \ref{assum:w_n_equal_v_n} hold. Then $\|h_{n,\epsilon}\|_{H^{0,1}(\Omega_T)}\to 0$ uniformly in $\epsilon$ as $n\to +\infty$.
    \begin{proof}
    Since $w_n \to 0$ in $L^2(\Omega_T)$ and $v_n \to 0$ in $H^{1,2}(\Omega_T)$, we have $e_n \to 0$ in $L^2(\Omega_T)$. Test \eqref{eq:PDE_for_h_{n,epsilon}} with $h_{n, \epsilon}$ and integrate over time and space,
    \begin{align*}
    &\int_0^T \int_\Omega \cG[h_{n, \epsilon}](t,x) h_{n, \epsilon}(t,x) \, dx \, dt + \int_0^T \langle \beta_{n,\epsilon}^t(z_{n, \epsilon}) - \beta_{n,\epsilon}^t(\hat{z}_{n, \epsilon}), h_{n, \epsilon} \rangle \, dt \\
    =&\int_0^T\int_\Omega e_n(t,x) h_{n, \epsilon}(t,x) \, dx \, dt.
    \end{align*}
    Since $\theta_\epsilon$ is increasing, we have
    \begin{align*}
    & \langle \beta_{n,\epsilon}^t(z_{n, \epsilon}) - \beta_{n,\epsilon}^t(\hat{z}_{n, \epsilon}), h_{n, \epsilon}-e_n \rangle \\
    =& \langle \beta_{n,\epsilon}^t(z_{n, \epsilon}) - \beta_{n,\epsilon}^t(\hat{z}_{n, \epsilon}), z_{n, \epsilon}-\hat{z}_{n,\epsilon} \rangle \\
    =& \int_\Omega G_n(t,x) (\theta_\epsilon(z_{n, \epsilon})-\theta_\epsilon(\hat{z}_{n, \epsilon})) (z_{n, \epsilon} - \hat{z}_{n, \epsilon}) \, dx + \int_{\Gamma_n^t} Q_n(t,x) (\theta_\epsilon(z_{n, \epsilon})-\theta_\epsilon(\hat{z}_{n, \epsilon})) (z_{n, \epsilon} - \hat{z}_{n, \epsilon}) \, dx \\
    \geq & 0.
    \end{align*}
    Hence,
    \begin{equation}
        \label{eq:estimation_for_rhs}
    \begin{aligned}
    &\int_0^T\int_\Omega \cG[h_{n, \epsilon}](t,x) h_{n, \epsilon}(t,x) \, dx \, dt\\
    =& - \int_0^T \langle \beta_{n,\epsilon}^t(z_{n, \epsilon}) - \beta_{n,\epsilon}^t(\hat{z}_{n, \epsilon}), h_{n, \epsilon} \rangle \, dt + \int_0^T\int_\Omega e_n(t,x) h_{n, \epsilon}(t,x) \, dx \, dt\\
    \leq& - \int_0^T \langle \beta_{n,\epsilon}^t(z_{n, \epsilon}) - \beta_{n,\epsilon}^t(\hat{z}_{n, \epsilon}), e_n \rangle \, dt + \int_0^T\int_\Omega e_n(t,x) h_{n, \epsilon}(t,x) \, dx \, dt\\
    \leq& K\int_0^T \left(\|e_n(t,\cdot)\|_{L^2(\Omega)} + \|e_n(t,\cdot)\|_{L^2(\Omega)}^{1/2} \|e_n(t,\cdot)\|_{H^1(\Omega)}^{1/2}\right) \, dt + \|e_n\|_{L^2(\Omega_T)} \|h_{n, \epsilon}\|_{L^2(\Omega_T)}\\
    \leq& K \left(\|e_n\|_{L^2(\Omega_T)} + \|e_n\|_{L^2(\Omega_T)}^{1/2} \|e_n\|_{H^{0,1}(\Omega_T)}^{1/2}\right) + \|e_n\|_{L^2(\Omega_T)} \|h_{n, \epsilon}\|_{L^2(\Omega_T)}\\
    \leq& K \|e_n\|_{L^2(\Omega_T)}^{1/2} + \|e_n\|_{L^2(\Omega_T)} \|h_{n, \epsilon}\|_{L^2(\Omega_T)},
    \end{aligned}
    \end{equation}
    where the last line follows from the uniform bound of $w_n$ and $v_n$ in $H^{0,1}(\Omega_T)$ and $\|e_n\|_{L^2(\Omega_T)}\to 0$. From \eqref{eq:operator_G} and \eqref{eq:bilinear_form}, since $h_{n, \epsilon} = 0$ on $\partial_p\Omega_T$, we have
    \begin{align*}
    \int_0^T\int_\Omega \cG[h_{n, \epsilon}](t,x) h_{n, \epsilon}(t,x) \, dx \, dt = & \int_0^T\int_\Omega \partial_t h_{n, \epsilon}(t,x) h_{n, \epsilon}(t,x) \, dx \, dt + \int_0^T a(t;h_{n, \epsilon},h_{n, \epsilon}) \, dt.
    \end{align*}
    Apply Lemma \ref{lem:exchange_of_differentiation_and_integration},
    \[
    \int_0^T\int_\Omega \partial_t h_{n, \epsilon}(t,x) h_{n, \epsilon}(t,x) \, dx \, dt = \frac{1}{2}\int_0^T \frac{d}{dt} \int_\Omega (h_{n, \epsilon}(t,x))^2 \, dx \, dt = \frac{1}{2}\|h_{n,\epsilon}(T,\cdot)\|^2_{L^2(\Omega)} \geq 0.
    \]
    From the coercivity in Assumption \ref{assum:assumptions_on_Theorem_8.2},
    \begin{align*}
    \int_0^T a(t;h_{n,\epsilon},h_{n,\epsilon}) \, dt \geq M \|h_{n,\epsilon}\|_{H^{0,1}(\Omega_T)}^2,
    \end{align*}
    where $M$ is a constant independent of $\epsilon$ and $n$. Hence, from \eqref{eq:estimation_for_rhs}, using Young's inequality,
    \begin{align*}
       & M \|h_{n,\epsilon}\|_{H^{0,1}(\Omega_T)}^2 \\
       \leq & K\|e_n\|_{L^2(\Omega_T)}^{1/2} + \|e_n\|_{L^2(\Omega_T)} \|h_{n, \epsilon}\|_{L^2(\Omega_T)} \\
       \leq & K\|e_n\|_{L^2(\Omega_T)}^{1/2} + \|e_n\|_{L^2(\Omega_T)} \|h_{n, \epsilon}\|_{H^{0,1}(\Omega_T)} \\
       \leq & K\|e_n\|_{L^2(\Omega_T)}^{1/2} + \frac{1}{2M}\|e_n\|_{L^2(\Omega_T)}^2 + \frac{M}{2}\|h_{n, \epsilon}\|_{H^{0,1}(\Omega_T)}^2,
    \end{align*}
    which gives
    \begin{align*}
    \|h_{n,\epsilon}\|_{H^{0,1}(\Omega_T)} \leq & \frac{2K}{M} \|e_n\|_{L^2(\Omega_T)}^{1/2} + \frac{1}{M^2}\|e_n\|_{L^2(\Omega_T)}^2.
    \end{align*}
    From $e_n \to 0$ in $L^2(\Omega_T)$, we have $h_{n,\epsilon}\to0$ uniformly in $\epsilon$ as $n \to +\infty$.

    \end{proof}
\end{lemma}
From \eqref{eq:bound_f_n_epsilon} and \eqref{eq:bound_hat_f_n_epsilon}, we have
\begin{align*}
    \|f_n-\hat{f}_n\|_{H^{0,1}(\Omega_T)} \leq & \|f_n - f_{n,\epsilon}\|_{H^{0,1}(\Omega_T)} + \|f_{n,\epsilon} - \hat{f}_{n,\epsilon}\|_{H^{0,1}(\Omega_T)} + \|\hat{f}_{n,\epsilon} - \hat{f}_n\|_{H^{0,1}(\Omega_T)} \\
    \leq & 2 C_n \epsilon^{1/2} + \|f_{n,\epsilon} - \hat{f}_{n,\epsilon}\|_{H^{0,1}(\Omega_T)}.
\end{align*}
For any $\delta>0$, there exists $N(\delta)$ such that for all $n \geq N(\delta)$ and all $\epsilon>0$, $\|f_{n,\epsilon}-\hat{f}_{n,\epsilon}\|_{H^{0,1}(\Omega_T)}<\delta$, so
\begin{align*}
    \|f_n-\hat{f}_n\|_{H^{0,1}(\Omega_T)} \leq 2 C_n \epsilon^{1/2} + \delta, \quad \forall n \geq N(\delta).
\end{align*}
Letting $\epsilon \to 0$, we obtain
\begin{align*}
    \|f_n-\hat{f}_n\|_{H^{0,1}(\Omega_T)} \leq \delta, \quad \forall n \geq N(\delta).
\end{align*}
Therefore, $\hat{f}_n - f_n \to 0$ strongly in $H^{0,1}(\Omega_T)$.
    
\end{proof}

\subsection{Proof of Theorem \ref{thm:g_n_to_u_in_unbounded_domain}} \label{app:proof_thm_5}
\begin{proof}
    We first prove the following lemma for the existence and uniqueness of $u^\infty$.
    \begin{lemma} \label{lem:existence_of_u_infty}
    Let Assumption \ref{assum:interior_estimates_for_u^k} holds. Suppose for any $k\in\bN$, Assumption \ref{assum:assumptions_on_Theorem_8.2} holds on $\Omega_T^k$. Then there exists a unique solution $u^\infty \in C(\overline{\Omega}_T^{\infty}) \cap W^{1,2}_{p, loc}(\Omega_T^{\infty})$ of \eqref{eq:unbounded_VI} for $1<p<+\infty$. Moreover, a single subsequence of $\{u^k\}_{k=1}^\infty$ converges to $u^\infty$ uniformly on $M$, weakly in $W_{p}^{1,2}(M)$ and strongly in $W_{p}^{0,1}(M)$ for any compact set $M \subset \Omega_T^{\infty}$.
    \begin{proof}
    The uniqueness of the solution can be attained from Lemma.A.1 in \cite{maGlobalClosedFormApproximation2019} under Assumption \ref{assum:interior_estimates_for_u^k}. For the existence, we follow the similar steps in \cite{maGlobalClosedFormApproximation2019} to construct $u^\infty$ from $\{u^k\}_{k=1}^\infty$. 
    
    Consider the following necessary condition for \eqref{eq:bounded_VI}:
        \begin{align*}
            \label{eq:necessary_condition_for_VI}
            \begin{cases}
                \cG[u^k](t,x) = I_{\{u^k=g\}} \cG[g](t,x), & \text{for } (t,x) \in \Omega_T^k, \\
                u^k(t,x) = g(t,x), & \text{for } (t,x) \in \partial_p\Omega_T^k,
            \end{cases}
        \end{align*}
    where $I_{\{u^k=g\}}$ is the indicator function of the set $\{u^k=g\}$.
    For any $0 < \xi < k$ and $1<p<+\infty$, from the $W_p^{1,2}$ interior estimate due to Assumption \ref{assum:interior_estimates_for_u^k},
    \begin{equation}
        \label{eq:interior_estimate_for_u^k}
    \|u^k\|_{W_p^{1,2}(\Omega_T^\xi)} \leq C_\xi \text{ where the constant } C_\xi \text{ only depends on } \xi \text{ but not on } k.
    \end{equation}
    Let $\xi = 1$ with $k \in \bZ^+ \cap (1, +\infty)$ in \eqref{eq:interior_estimate_for_u^k}. Since $W_p^{1,2}(\Omega_T^1)$ is reflexive, there is a subsequence $\{u^{k_l^{(1)}}\}_{l \in \bZ^+}$ of $\{u^k\}_{k \in \bZ^+ \cap (1, +\infty)}$ that converges weakly to a limit $u^{(1)}$ in $W_p^{1,2}(\Omega_T^1)$. By Rellich-Kondrachov theorem, there exist two compact embeddings: 
    \[W^{2,p}(\Omega^1) \subset\subset W^{1,p}(\Omega^1) \subset\subset L^p(\Omega^1).\]
    From $u^{k_l^{(1)}} \in W^{1,2}_p(\Omega_T^1)$, we know $u^{k_l^{(1)}} \in L^p(0,T; W^{2,p}(\Omega^1))$ and $\partial_t u^{k_l^{(1)}} \in L^p(\Omega_T^1)$. By the Aubin-Lions Theorem, there is a further subsequence that converges to $u^{(1)}$ strongly in $L^p(0,T;W^{1,p}(\Omega^1))$, i.e., $W_p^{0,1}(\Omega_T^1).$ Moreover, by the compact embedding into $C^0(\overline{\Omega_T^1})$ from Rellich-Kondrachov theorem when $p > d$, there is again a further subsequence that converges to $u^{(1)}$ in $C^0(\overline{\Omega_T^1})$. Note we took subsequence for three different convergences: weak convergence, strong convergence and uniform convergence. For simplicity, we still denote the subsequence as $\{u^{k_l^{(1)}}\}_{l \in \bZ^+}$. We conclude
    \[
    u^{k_l^{(1)}} \xrightarrow{\text{$l \to \infty$}} u^{(1)} \text{ weakly in } W_p^{1,2}(\Omega_T^1), \text{ strongly in } W_p^{0,1}(\Omega_T^1) \text{ and strongly in } C^0(\overline{\Omega_T^1}).
    \]
    
    Let $\xi = 2$ with $k \in \{k_l^{(1)}\}_{l\in \bZ^+ \cap (2, +\infty)}$ in \eqref{eq:interior_estimate_for_u^k} so that $\xi < k$, similarly as above, by weak compactness, Sobolev embedding and the Aubin-Lions Theorem, there is a subsequence $\{u^{k_l^{(2)}}\}_{l \in \bZ^+}$ of $\{u^{k_l^{(1)}}\}_{l\in \bZ^+ \cap (2, +\infty)}$ such that 
    \[
    u^{k_l^{(2)}} \xrightarrow{\text{$l \to \infty$}} u^{(2)} \text{ weakly in } W_p^{1,2}(\Omega_T^2), \text{ strongly in } W_p^{0,1}(\Omega_T^2) \text{ and strongly in } C^0(\overline{\Omega_T^2}).
    \]
    Moreover, due to the subsequence property, we have 
    \[
    u^{(1)} = u^{(2)} \text{ in } \Omega_T^1.
    \]
    
    Denote $k_l^{(0)} = l$. By induction from above, for any $m \in \bZ^+$, we can construct a subsequence $\{u^{k_l^{(m)}}\}_{l \in \bZ^+}$ of $\{u^{k_l^{(m-1)}}\}_{l \in \bZ^+}$ in $\Omega_T^m$ such that 
    \[
    u^{k_l^{(m)}} \xrightarrow{\text{$l \to \infty$}} u^{(m)} \text{ weakly in } W_p^{1,2}(\Omega_T^m), \text{ strongly in } W_p^{0,1}(\Omega_T^m) \text{ and strongly in } C^0(\overline{\Omega_T^m}).
    \]
    Moreover, 
    \[
    u^{(m)} = u^{(j)} \quad \text{in } \Omega_T^j, \quad 1 \leq j \leq m-1.
    \]
    
    Define $u^\infty := u^{(m)}$ in $\Omega_T^m$ for any $m \in \bZ^+$. Write the sequences in a matrix:
    \begin{align*}
        & u^1, u^2, u^3, \cdots \\ & u^{k^{(1)}_1}, u^{k^{(1)}_2}, u^{k^{(1)}_3}, \cdots \to u^{(1)} \\ & u^{k^{(2)}_1}, u^{k^{(2)}_2}, u^{k^{(2)}_3}, \cdots \to u^{(2)} \\ \vdots
    \end{align*}
    
    Note $k_m^{(m)} > k_{m-1}^{(m-1)}$ by previous construction. Therefore, it is valid to consider the diagonal sequence $\{u^{k^{(m)}_m}\}_{m \in \bZ^+}$. For any $N \in \bZ^+$, since $\{u^{k^{(m)}_m}\}_{m \in \bZ^+ \cap [N,+\infty)}$ is a subsequence of $\{u^{k^{(N)}_m}\}_{m \in \bZ^+ \cap [N,+\infty)}$, we derive that 
    \[
    u^{k^{(m)}_m} \xrightarrow{\text{$m \to \infty$}} u^{(N)} = u^\infty \text{ weakly in } W^{1,2}_p(\Omega_T^N), \text{ strongly in } W^{0,1}_p(\Omega_T^N) \text{ and } C^0(\overline{\Omega_T^N}).
    \]
    
    Note $u^{k_m^{(m)}}$ satisfies \eqref{eq:bounded_VI} with $k=k_m^{(m)}$. For $N \in \bZ^+$ s.t. $N \leq m < k_m^{(m)}$, the following  system holds:
    \begin{align}
    \label{eq:VI_u^{k_m^{(m)}}}
        \begin{cases}
            \min\{\cG[u^{k_m^{(m)}}](t,x), u^{k_m^{(m)}}(t,x)-g(t,x)\} = 0, & \text{for } (t,x) \in \Omega_T^N, \\
            u^{k_m^{(m)}}(0,x) = g(0,x), & \text{for } x \in \Omega^N.
        \end{cases}
    \end{align}
    Following the definition of $u^\infty$ from above, we would prove 
    \begin{align}
    \label{eq:VI_u^{k_m^{(m)}}_limit}
        \forall N \in \bZ^+, \quad
        \begin{cases}
            \min\{\cG[u^\infty](t,x), u^\infty(t,x)-g(t,x)\} = 0, & \text{for } (t,x) \in \Omega_T^N, \\
            u^\infty(0,x) = g(0,x), & \text{for } x \in \Omega^N.
        \end{cases}
    \end{align}
    Similarly to \ref{app:proof_thm_2}, there is a series of variational inequalities and a "limit" variational inequality. Better than Theorem \ref{thm:hat_f_n_converges_to_u}, now boundary conditions, obstacles and source terms are all the same, so the admissible sets are the same. The weak form of \eqref{eq:VI_u^{k_m^{(m)}}} is 
    \[
    u^{k_m^{(m)}} \in K, \quad \int_{\Omega^N} \cG[u^{k_m^{(m)}}] (\phi-u^{k_m^{(m)}}) \, dx \geq 0, \quad \text{ for a.a. } t \in (0,T), \quad \forall \phi \in K,
    \]
    where $K = \{\phi \ | \ \phi \in H^1(\Omega_T^N), \phi(0,x) = g(0,x) \text{ on } \Omega^N, \phi \geq g \text{ a.e.}\}.$ The weak form of \eqref{eq:VI_u^{k_m^{(m)}}_limit} is
    \[
    u^\infty \in K, \quad \int_{\Omega^N} \cG[u^\infty] (\phi-u^\infty) \, dx \geq 0, \quad \text{ for a.a. } t \in (0,T), \quad \forall \phi \in K.
    \]
    Given \eqref{eq:VI_u^{k_m^{(m)}}} holds, letting $m \to +\infty$ in the admissible set, we know $u^\infty \in K$. Moreover, by following the similar integration by parts as in the proof of Theorem \ref{thm:hat_f_n_converges_to_u}, we know $u^\infty$ satisfies the weak form of \eqref{eq:VI_u^{k_m^{(m)}}_limit}. Since $N$ is arbitrary and the solution to \eqref{eq:unbounded_VI} is unique, $u^\infty$ is the solution to \eqref{eq:unbounded_VI}. From previous convergence results, $u^\infty \in C(\overline{\Omega}_T^{\infty}) \cap W^{1,2}_{p, loc}(\Omega_T^{\infty})$ for $1<p<+\infty$.

    \end{proof}

    \end{lemma}

    Next, we prove the convergence of the neural networks. Under the given conditions, from Theorem \ref{thm:f_n_converges_to_u}, there exists a sequence $\{f_n^k\}_{n=1}^\infty$ such that $J(f_n^k)\to0$ and $f_n^k$ converges to $u^{k}$ strongly in $H^{0,1}(\Omega_T^{k})$. Hence, there exists $N(k)$ such that for any $n \geq N(k)$,
    \[\|f_{n}^k - u^k\|_{H^{0,1}(\Omega_T^k)} < 1/k.\]
    Set $\phi_k := f_{N(k)}^k$. Then 
    \[
    \|\phi_k - u^k\|_{H^{0,1}(\Omega_T^k)} < 1/k.
    \]
    
    From Lemma \ref{lem:existence_of_u_infty}, there is a single sequence $\{u^{k_m^{(m)}}\}_{m=1}^\infty$ that converges to $u^\infty$ in $H^{0,1}(\Omega_T^N)$ for any $N \in \bZ^+$.
    Set $g_m := \phi_{k_m^{(m)}}$. Then
    \[
    \|g_m - u^{k_m^{(m)}}\|_{H^{0,1}\left(\Omega_T^{k_m^{(m)}}\right)} < 1/k_m^{(m)}.
    \]
    For a fixed compact set $M \in \Omega_T^{\infty}$, there exists $m_0 \in \bZ^+$ such that $M \subset \Omega_T^{k_{m_0}^{(m_0)}}$.
    For any $m \geq m_0$, 
    \begin{align*}
    \|g_m - u^\infty\|_{H^{0,1}\left(\Omega_T^{k_{m_0}^{(m_0)}}\right)} \leq & \|g_m - u^{k_{m}^{(m)}}\|_{H^{0,1}\left(\Omega_T^{k_{m_0}^{(m_0)}}\right)} + \|u^{k_{m}^{(m)}} - u^\infty\|_{H^{0,1}\left(\Omega_T^{k_{m_0}^{(m_0)}}\right)} \\
    \leq & \|g_m - u^{k_{m}^{(m)}}\|_{H^{0,1}\left(\Omega_T^{k_{m}^{(m)}}\right)} + \|u^{k_{m}^{(m)}} - u^\infty\|_{H^{0,1}\left(\Omega_T^{k_{m_0}^{(m_0)}}\right)} \\
    < & 1/k_{m}^{(m)} + \|u^{k_{m}^{(m)}} - u^\infty\|_{H^{0,1}\left(\Omega_T^{k_{m_0}^{(m_0)}}\right)}.
    \end{align*}
    For any $\epsilon > 0$, both of the terms can be controlled by $\epsilon$ as follows.
    Since $k_m^{(m)} \to +\infty$, there exists $Z_1(\epsilon)$ such that for any $m \geq Z_1(\epsilon)$,
    \[
    1/k_{m}^{(m)} < \epsilon/2.
    \]
    Since $u^{k_m^{(m)}}\to u^\infty$ in $H^{0,1}(\Omega_T^N)$ for any $N\in\bZ^+$, there exists $Z_2(\epsilon, M)$ such that for any $m \geq Z_2(\epsilon, M)$,
    \[
    \|u^{k_{m}^{(m)}} - u^\infty\|_{H^{0,1}\left(\Omega_T^{k_{m_0}^{(m_0)}}\right)} < \epsilon/2.
    \]
    Therefore, for any $\epsilon > 0$, there exists $Z(\epsilon,M) = \max\{m_0, Z_2(\epsilon, M), Z_1(\epsilon)\}$ such that for any $m \geq Z(\epsilon,M)$,
    \[
    \|g_m - u^\infty\|_{H^{0,1}(M)} \leq \|g_m - u^\infty\|_{H^{0,1}\left(\Omega_T^{k_{m_0}^{(m_0)}}\right)} < \epsilon.
    \]
    Since $Z(\epsilon,M)$ depends on $M$, a convergence in $H^{0,1}(\Omega_T^{\infty})$ generally does not exist.
    
\end{proof}

\subsection{Proof of Corollary \ref{cor:g_n_to_u_a.a. time}} \label{app:proof_cor_1}
\begin{proof}
    From Theorem \ref{thm:g_n_to_u_in_unbounded_domain}, $\forall k \in \bZ^+$, $g_n\to u^\infty$ in $H^{0,1}(\Omega_T^k)$. Denote $l_n^k(t) := \|g_n(t, \cdot) - u^\infty(t, \cdot)\|_{H^1(\Omega^k)}$. Then $l_n^k(t)\to0$ in $L^2((0, T))$. Therefore, a subsequence of $\{l_n^k(t)\}_{n=1}^\infty$ converges to $0$ a.e. in $(0, T)$. Denote such a subsequence as $\{l_{n_m^{(k)}}^k(t)\}_{m=1}^\infty$.
    
    Accordingly, the subsequence $\{g_{n_m^{(k)}}(t;x)\}_{m=1}^\infty$ converges to $u^\infty(t;x)$ in $H^1(\Omega^k)$ at almost any $t \in (0,T)$. However, such a choice of subsequence $n_m^{(k)}$ depends on $k$.
    
    Since $g_n$ also converges to $u^\infty$ in $H^{0,1}(\Omega_T^{k+1})$, we know its subsequence $\{g_{n_m^{(k)}}(t,x)\}_{m=1}^\infty$ also converges to $u^\infty(t,x)$ in $H^{0,1}(\Omega_T^{k+1})$. Similarly to the above, we can extract a further subsequence from it such that this sub-subsequence converges to $u^\infty(t;x)$ in $H^1(\Omega^{k+1})$ at almost any $t \in (0,T)$. Denote such a subsequence as $\{g_{n_{m}^{(k+1)}}(t,x)\}_{m=1}^\infty$. 
    
    Following the similar process, by induction, $\forall k \in \bZ^+$, we can construct a subsequence $\{g_{n_m^{(k)}}(t;x)\}_{m=1}^\infty$ that converges to $u^\infty(t;x)$ in $\Omega^k$ at almost any $t \in (0,T)$. Moreover, $\{n_m^{(k+1)}\}_{m=1}^\infty$ is a subsequence of $\{n_m^{(k)}\}_{m=1}^\infty$.
    
    Write the indices and spaces like a matrix:
    \begin{align*}
        & n_1^{(1)}, n_2^{(1)}, n_3^{(1)}, \cdots \quad \Omega^1 \\
        & n_1^{(2)}, n_2^{(2)}, n_3^{(2)}, \cdots \quad \Omega^2 \\
        & n_1^{(3)}, n_2^{(3)}, n_3^{(3)}, \cdots \quad \Omega^3 \\ 
        \vdots
    \end{align*}
    
    Define $v_m := g_{n_m^{(m)}}$. Since the measure of a countable union of zero-measure sets is zero, we conclude $\forall k \in \bZ^+$, $v_m(t;x)\to u^\infty(t;x)$ in $H^1(\Omega^k)$ at almost any $t \in (0,T)$.
        
\end{proof}

\subsection{Proof of Proposition \ref{prop:convergence_after_variable_transform}} \label{app:proof_prop_2}
\begin{proof}
    From Theorem \ref{thm:g_n_to_u_in_unbounded_domain}, there is a sequence of neural networks $\{g_n\}_{n=1}^\infty$ that converges to $v$ in $H^{0,1}(M')$ for any compact set $M' \subset Q_z$.

    Define $h_n(t, y) := g_n(\tau, z)$. For any compact set $M \subset Q_y$, denote its image as $M' \subset Q_z$. Since a continuous image of a compact set is compact, and a homeomorphism in particular is a continuous bijection with continuous inverse, we know $M'$ is also compact. By \eqref{def:variable_transform},
    \[
    \int_{M} \left(h_n(t,y) - \tilde{V}(t,y)\right)^2 \ dy dt = -\frac{2}{\theta^2} \int_{M'} e^z \left(g_n(\tau,z) - v(\tau,z)\right)^2 \ dz d\tau.
    \]
    Since $M'$ is compact, we know $e^z$ is bounded in $M'$, so $h_n(t,y)$ strongly converge to $\tilde{V}(t,y)$ in $L^2(M)$. Similarly, note that $D_y h_n(t,y) = e^z D_z g_n(\tau, z)$ and $D_y \tilde{V}(t, y) = e^z D_z v(t, z)$,
    \[
    \int_{M} \left(D_y h_n(t,y) - D_y \tilde{V}(t,y)\right)^2 \ dy dt = -\frac{2}{\theta^2} \int_{M'} e^{3z} \left(D_z g_n(\tau,z) - D_z v(\tau,z)\right)^2 \ dz d\tau.
    \]
    By similar arguments, $D_y h_n(t,y)$ strongly converge to $D_y \tilde{V}(t,y)$ in $L^2(M)$. We  conclude that $\{h_n(t,y)\}_{n=1}^\infty$ strongly converges to $\tilde{V}(t,y)$ in $H^{0,1}(M)$ and such a sequence is independent of $M$.
    
\end{proof}

\subsection{Proof of Theorem \ref{thm:uniform_convergence_to_primal_value}} \label{app:proof_thm_7}
\begin{proof}
    We first prove the uniform convergence. From Corollary \ref{cor:convergence_after_variable_transform_a.a. time}, for almost any $t \in [0,T]$, $\{w_n(t;y)\}_{n=1}^\infty$ converges to $\tilde{V}(t;y)$ in $C^{0, \frac{1}{2}}(M_y(M_x))$. Therefore, for any $\epsilon > 0$, there exists $N(\epsilon)$ such that for all $n > N(\epsilon)$, 
    \[
    \sup_{y \in M_y(M_x)}|w_n(t;y) - \tilde{V}(t;y)| < \epsilon.
    \]
    
    Since $y_x^t$ and $y_{x,n}^t$ belong to $M_y(M_x)$, by \eqref{eq:dual_transform} and \eqref{def:V_n}, the following relations hold:
    \begin{align*}
        V(t;x) & = \tilde{V}(t;y_x^t) + xy_x^t, \\
        V_{n}(t;x) & \leq w_n(t;y_x^t) + xy_x^t, \\
        V(t;x) & \leq \tilde{V}(t;y_{x,{n}}^t) + xy_{x,{n}}^t, \\
        V_{n}(t;x) & = w_n(t;y_{x,{n}}^t) + xy_{x,{n}}^t.
    \end{align*}
    $\forall n > N(\epsilon)$, taking the difference between the first two relations and the last two relations, we have
    \[
    -\epsilon < w_n(t;y_{x,{n}}^t) - \tilde{V}(t;y_{x,{n}}^t) \leq V_{n}(t;x) - V(t;x) \leq w_n(t;y_x^t) - \tilde{V}(t;y_x^t) < \epsilon, \ \forall x \in M_x.
    \]
    Therefore, $\{V_{n}(t;x)\}_{k=1}^\infty$ converges to $V(t;x)$ uniformly on $M_x$ for almost any $t \in [0,T]$.

    Next, we prove the convergence in $L^2(0,T; L^\infty(M_x))$.
    Denote
    \[
    \phi_x^t(y):= \tilde{V}(t,y)+xy, \quad \phi^t_{n,x}(y):= w_n(t,y)+xy.
    \]
    Then we have
    \[
    V(t,x)=\inf_{y\in M_y(M_x)} \phi^t_x(y), \quad V_n(t,x)=\inf_{y\in M_y(M_x)} \phi^t_{n,x}(y).
    \]
    Denote $\Delta:=\sup_{y\in M_y(M_x)} \left|\phi^t_x(y) - \phi^t_{n,x}(y)\right|<\infty$. Then we have
    \[
    \phi^t_x(y) - \Delta \leq \phi^t_{n,x}(y) \leq \phi^t_x(y) + \Delta, \quad \forall y \in M_y(M_x).
    \]
    Therefore,
    \begin{align*}
        -\Delta \leq \inf_{y\in M_y(M_x)} (\phi^t_{n,x}(y)-\phi^t_x(y)) \leq \inf_{y\in M_y(M_x)} \phi^t_{n,x}(y) - \inf_{y\in M_y(M_x)} \phi^t_x(y).
    \end{align*}
    \begin{align*}
        \Delta \geq \sup_{y\in M_y(M_x)} (\phi^t_{n,x}(y)-\phi^t_x(y)) \geq \inf_{y\in M_y(M_x)} \phi^t_{n,x}(y) - \inf_{y\in M_y(M_x)} \phi^t_x(y).
    \end{align*}
    By using the above transformations, we obtain
    \begin{align*}
        \left|\inf_{y\in M_y(M_x)} \phi^t_{n,x}(y) - \inf_{y\in M_y(M_x)} \phi^t_x(y)\right| & \leq \sup_{y\in M_y(M_x)} \left|\phi^t_x(y) - \phi^t_{n,x}(y)\right| \\
        & = \sup_{y\in M_y(M_x)} |\tilde{V}(t,y)-w_n(t,y)|.
    \end{align*}
    Define $\Phi_n(t):=\sup_{y\in M_y(M_x)} |\tilde{V}(t,y)-w_n(t,y)|$. Then the above implies
    \[
    \left|V_n(t,x) - V(t,x)\right| \leq \Phi_n(t), \quad \forall (t,x) \in (0,T)\times M_x.
    \]
    Denote $a_n(t,y):=\tilde{V}(t,y)-w_n(t,y)$. For almost any $t \in (0,T)$, $a_n(t,\cdot)\in H^1(M_y(M_x))$. Since in one dimension, the Sobolev embedding theorem gives us $H^1(M_y(M_x)) \hookrightarrow C^0(M_y(M_x))$ and $C^0(M_y(M_x)) \subset L^\infty(M_y(M_x))$, we have
    \[
    \Phi_n(t) = \sup_{y\in M_y(M_x)} \left|a_n(t,y)\right| = \|a_n(t,\cdot)\|_{L^\infty(M_y(M_x))} \leq K \|a_n(t,\cdot)\|_{H^1(M_y(M_x))},
    \]
    where $K$ only depends on $M_y(M_x)$.
    Then we obtain
    \begin{align*}
        \|\Phi_n\|_{L^2(0,T)}^2 & \leq K^2 \int_0^T \|a_n(t,\cdot)\|_{H^1(M_y(M_x))}^2 \, dt \\
        & = K^2 \|\tilde{V}-w_n\|_{H^{0,1}((0,T)\times M_y(M_x))}^2 \to 0.
    \end{align*}
    Therefore, $\Phi_n \to 0$ in $L^2(0,T)$.
    We have the estimate
    \begin{align*}
        \sup_{x \in M_x} \left|V_n(t,x) - V(t,x)\right| & \leq \Phi_n(t).
    \end{align*}
    Hence, $\sup_{x \in M_x} \left|V_n(t,x) - V(t,x)\right| \to 0$ in $L^2(0,T)$. This implies that $V_n(t,x) \to V(t,x)$ in $L^2(0,T; L^\infty(M_x))$.

\end{proof}

\section{Conclusions} \label{sec:conclusions}
In this paper, we provide a rigorous convergence proof of deep neural networks to the solution of linear parabolic variational inequalities, and extend it to nonlinear operators using dual method. We design the loss function from trace theorems, prove the existence of neural networks that drive the loss function to zero and converge to the true solution in a Sobolev space. The bounded domain results are then extended to unbounded domain. We apply the results on the mixed optimal stopping and control problem, and use dual method  to prove the convergence to primal value function. We also provide numerical examples to show the performance of the neural network method under power and non-HARA utilities, and compare it with other methods. We test the approach for high-dimensional American put option pricing and show that the neural network method is accurate and efficient. For general nonlinear variational inequalities, it is difficult to find a uniform approach to transform it into linear operators, and our approach may not be applicable. We leave this for future work.

\bigskip\noindent
{\bf Acknowledgment}.  The authors thank Yufei Zhang for the discussion on the inverse trace theorem which helps to improve the  formulation of the loss function.


\begin{thebibliography}{34}
\providecommand{\natexlab}[1]{#1}
\providecommand{\url}[1]{\texttt{#1}}
\expandafter\ifx\csname urlstyle\endcsname\relax
  \providecommand{\doi}[1]{doi: #1}\else
  \providecommand{\doi}{doi: \begingroup \urlstyle{rm}\Url}\fi

\bibitem[Adams and Lenhart(2002)]{adamsOptimalControlObstacle2002}
D.~R.\ Adams and S.\ Lenhart.
\newblock Optimal control of the obstacle for a parabolic variational
  inequality.
\newblock \emph{Journal of Mathematical Analysis and Applications},
  268\penalty0 (2):\penalty0 602--614, 2002.

\bibitem[Alphonse et~al.(2024)Alphonse, Hinterm{\"u}ller, Kister, Lun, and
  Sirotenko]{alphonseNeuralNetworkApproach2024}
A.\ Alphonse, M.\ Hinterm{\"u}ller, A.\ Kister, C.~H.\ Lun, and C.\ Sirotenko.
\newblock A neural network approach to learning solutions of a class of
  elliptic variational inequalities, 2024, arXiv:2411.18565.

\bibitem[Aubin and Frankowska(2009)]{aubinSetValuedAnalysis2009}
J.~P.\ Aubin and H.\ Frankowska.
\newblock \emph{Set-{{Valued Analysis}}}.
\newblock Birkh{\"a}user, 2009.

\bibitem[Barbu(1993)]{barbuAnalysisControlNonlinear1993}
V.\ Barbu.
\newblock \emph{Analysis and {{Control}} of {{Nonlinear Infinite Dimensional
  Systems}}}.
\newblock Academic press, 1993.

\bibitem[Becker et~al.(2019)Becker, Cheridito, and
  Jentzen]{beckerDeepOptimalStopping2019}
S.\ Becker, P.\ Cheridito, and A.\ Jentzen.
\newblock Deep optimal stopping.
\newblock \emph{The Journal of Machine Learning Research}, 20\penalty0
  (1):\penalty0 2712--2736, 2019.

\bibitem[Becker et~al.(2021)Becker, Cheridito, Jentzen, and
  Welti]{beckerSolvingHighdimensionalOptimal2021}
S.\ Becker, P.\ Cheridito, A.\ Jentzen, and T.\ Welti.
\newblock Solving high-dimensional optimal stopping problems using deep
  learning.
\newblock \emph{European Journal of Applied Mathematics}, 32\penalty0
  (3):\penalty0 470--514, 2021.

\bibitem[Bensoussan and
  Lions(2011)]{bensoussanApplicationsVariationalInequalities2011}
A.\ Bensoussan and J.-L.\ Lions, editors.
\newblock \emph{Applications of {{Variational Inequalities}} in {{Stochastic
  Control}}}.
\newblock North-Holland Pub. Co, 2011.

\bibitem[Boyer and Fabrie(2013)]{boyerMathematicalToolsStudy2013}
F.\ Boyer and P.\ Fabrie.
\newblock \emph{Mathematical {{Tools}} for the {{Study}} of the
  {{Incompressible Navier-Stokes Equations}} and {{Related Models}}}.
\newblock Springer, 2013.

\bibitem[Br{\'e}zis(2011)]{brezisFunctionalAnalysisSobolev2011}
H.\ Br{\'e}zis.
\newblock \emph{Functional {{Analysis}}, {{Sobolev Spaces}} and {{Partial
  Differential Equations}}}.
\newblock Springer New York, 2011.

\bibitem[Broadie and Detemple(1997)]{broadieValuationAmericanOptions1997}
M.\ Broadie and J.\ Detemple.
\newblock The valuation of {{American}} options on multiple assets.
\newblock \emph{Mathematical Finance}, 7\penalty0 (3):\penalty0 241--286, 1997.

\bibitem[Dautov(2021)]{dautovPenaltyMethodsOneSided2021}
R.~Z.\ Dautov.
\newblock Penalty {{Methods}} for {{One-Sided Parabolic Problems}} with
  {{Piecewise Smooth Obstacles}}.
\newblock \emph{Lobachevskii Journal of Mathematics}, 42\penalty0 (7):\penalty0
  1643--1651, 2021.

\bibitem[Demengel and Demengel(2012)]{demengelFunctionalSpacesTheory2012}
F.\ Demengel and G.\ Demengel.
\newblock \emph{Functional {{Spaces}} for the {{Theory}} of {{Elliptic Partial
  Differential Equations}}}.
\newblock Springer London, 2012.

\bibitem[Evans(2010)]{evansPartialDifferentialEquations2010}
L.~C.\ Evans.
\newblock \emph{Partial {{Differential Equations}}}.
\newblock American mathematical society, 2nd ed edition, 2010.

\bibitem[Friedman(1982)]{friedmanVariationalPrinciplesFreeboundary1982}
A.\ Friedman.
\newblock \emph{Variational {{Principles}} and {{Free-Boundary Problems}}}.
\newblock Wiley, 1982.

\bibitem[Grisvard(1985)]{grisvardEllipticProblemsNonsmooth1985}
P.\ Grisvard.
\newblock \emph{Elliptic {{Problems}} in {{Nonsmooth Domains}}}.
\newblock Pitman, 1985.

\bibitem[Hornik(1991)]{hornikApproximationCapabilitiesMultilayer1991}
K.\ Hornik.
\newblock Approximation capabilities of multilayer feedforward networks.
\newblock \emph{Neural Networks}, 4\penalty0 (2):\penalty0 251--257, 1991.

\bibitem[Hur{\'e} et~al.(2020)Hur{\'e}, Pham, and
  Warin]{hureDeepBackwardSchemes2020}
C.\ Hur{\'e}, H.\ Pham, and X.\ Warin.
\newblock Deep backward schemes for high-dimensional nonlinear {{PDEs}}.
\newblock \emph{Mathematics of Computation}, 89\penalty0 (324):\penalty0
  1547--1579, 2020.

\bibitem[Ito et~al.(2021)Ito, Reisinger, and
  Zhang]{itoNeuralNetworkBasedPolicy2021}
K.\ Ito, C.\ Reisinger, and Y.\ Zhang.
\newblock A neural network-based policy iteration algorithm with global h
  2-superlinear convergence for stochastic games on domains.
\newblock \emph{Foundations of Computational Mathematics}, 21\penalty0
  (2):\penalty0 331--374, 2021.

\bibitem[Jaillet et~al.(1990)Jaillet, Lamberton, and
  Lapeyre]{jailletVariationalInequalitiesPricing1990a}
P.\ Jaillet, D.\ Lamberton, and B.\ Lapeyre.
\newblock Variational inequalities and the pricing of {{American}} options.
\newblock \emph{Acta Applicandae Mathematicae}, 21\penalty0 (3):\penalty0
  263--289, 1990.

\bibitem[Jang et~al.(2024)Jang, Xu, and
  Zheng]{jangOptimalInvestmentHeterogeneous2024}
H.~J.\ Jang, Z.~Q.\ Xu, and H.\ Zheng.
\newblock Optimal investment, heterogeneous consumption, and best time for
  retirement.
\newblock \emph{Operations Research}, 72\penalty0 (2):\penalty0 832--847, 2024.

\bibitem[Kinderlehrer and
  Stampacchia(1980)]{kinderlehrerIntroductionVariationalInequalities1980}
D.\ Kinderlehrer and G.\ Stampacchia.
\newblock \emph{An {{Introduction}} to {{Variational Inequalities}} and {{Their
  Applications}}}.
\newblock Academic Press, 1980.

\bibitem[Leoni(2009)]{leoniFirstCourseSobolev2009}
G.\ Leoni.
\newblock \emph{A {{First Course}} in {{Sobolev Spaces}}}.
\newblock American mathematical society, 2009.

\bibitem[Lions and
  Magenes(1972{\natexlab{a}})]{lionsNonHomogeneousBoundaryValue1972}
J.~L.\ Lions and E.\ Magenes.
\newblock \emph{Non-{{Homogeneous Boundary Value Problems}} and
  {{Applications}}: {{Volume I}}}.
\newblock Springer Berlin Heidelberg, 1972{\natexlab{a}}.

\bibitem[Lions and
  Magenes(1972{\natexlab{b}})]{lionsNonHomogeneousBoundaryValue1972a}
J.~L.\ Lions and E.\ Magenes.
\newblock \emph{Non-{{Homogeneous Boundary Value Problems}} and
  {{Applications}}: {{Volume II}}}.
\newblock Springer Berlin Heidelberg, 1972{\natexlab{b}}.

\bibitem[Longstaff and Schwartz(2001)]{longstaffValuingAmericanOptions2001}
F.~A.\ Longstaff and E.~S.\ Schwartz.
\newblock Valuing {{American}} options by simulation: A simple least-squares
  approach.
\newblock \emph{Review of Financial Studies}, 14\penalty0 (1):\penalty0
  113--147, 2001.

\bibitem[Lunardi(1995)]{lunardiAnalyticSemigroupsOptimal1995}
A.\ Lunardi.
\newblock \emph{Analytic {{Semigroups}} and {{Optimal Regularity}} in
  {{Parabolic Problems}}}.
\newblock Birkh{\"a}user, 1995.

\bibitem[Ma et~al.(2019)Ma, Xing, and
  Zheng]{maGlobalClosedFormApproximation2019}
J.\ Ma, J.\ Xing, and H.\ Zheng.
\newblock Global closed-form approximation of free boundary for optimal
  investment stopping problems.
\newblock \emph{SIAM Journal on Control and Optimization}, 57\penalty0
  (3):\penalty0 2092--2121, 2019.

\bibitem[Messerschmidt(2019)]{messerschmidtPointwiseLipschitzSelection2019}
M.\ Messerschmidt.
\newblock A pointwise lipschitz selection theorem.
\newblock \emph{Set-Valued and Variational Analysis}, 27\penalty0 (1):\penalty0
  223--240, 2019.

\bibitem[Peskir and Shiryaev(2006)]{peskirOptimalStoppingFreeboundary2006}
G.\ Peskir and A.~N.\ Shiryaev.
\newblock \emph{Optimal Stopping and Free-Boundary Problems}.
\newblock Birkh{\"a}user Verlag, 2006.

\bibitem[Rudin(1976)]{rudinPrinciplesMathematicalAnalysis1976}
W.\ Rudin.
\newblock \emph{Principles of {{Mathematical Analysis}}}.
\newblock McGraw-Hill, 3rd edition, 1976.

\bibitem[Sirignano and Spiliopoulos(2018)]{sirignanoDGMDeepLearning2018}
J.\ Sirignano and K.\ Spiliopoulos.
\newblock {{DGM}}: {{A}} deep learning algorithm for solving partial
  differential equations.
\newblock \emph{Journal of Computational Physics}, 375:\penalty0 1339--1364,
  2018.

\bibitem[Taylor(2023)]{taylorPartialDifferentialEquations2023}
M.~E.\ Taylor.
\newblock \emph{Partial {{Differential Equations I}}: {{Basic Theory}}}.
\newblock Springer, 3rd ed. 2023 edition, 2023.

\bibitem[Wang and Perdikaris(2021)]{wangDeepLearningFree2021a}
S.\ Wang and P.\ Perdikaris.
\newblock Deep learning of free boundary and {{Stefan}} problems.
\newblock \emph{Journal of Computational Physics}, 428:\penalty0 109914, 2021.

\bibitem[Whitney(1992)]{Whitney1992}
H.\ Whitney.
\newblock Analytic extensions of differentiable functions defined in closed
  sets.
\newblock In J.\ Eells and D.\ Toledo, editors, \emph{Hassler Whitney Collected
  Papers}, pages 228--254. Birkh{\"a}user Boston, 1992.

\end{thebibliography}

\end{document}